\documentclass[
	amsmath,
	amssymb,
	a4paper,
	aip,		% society option (aps, aip)
	jcp,		% society's journal style (prl, apl, ..)
	reprint, twocolumn  %reprint,	% layout option (preprint, reprint, twocolumn)
	fleqn,
	showpacs,
	floatfix
]{revtex4-1}

\usepackage[T1]{fontenc}
\usepackage{lmodern}
\usepackage{microtype}
\usepackage{color}
\usepackage{tabularx}
\usepackage{booktabs}
\usepackage{multirow}
\usepackage{graphicx} % Include figure files
\usepackage{dcolumn} % Align table columns on decimal point
\usepackage{bm} % bold math
\usepackage[version=3]{mhchem}
\usepackage{stackrel}

\usepackage{verbatim}   % useful for program listings
\usepackage{float}
\usepackage{amsfonts} %helps out with particular math fonts, e.g. Z and R
\usepackage[utf8]{inputenc}
\usepackage[english]{babel}
\usepackage{calc}

\usepackage{geometry}
\addto\captionsenglish{}

\usepackage{amsmath}                                                                                      
\usepackage{epsfig}                                                                                       
\usepackage{amssymb}

\usepackage{mathrsfs}
\usepackage{color}    
\definecolor{darkblue}{rgb}{0,0,0.6}
\definecolor{darkred}{rgb}{0.6,0,0}
\definecolor{darkgreen}{rgb}{0,0.6,0}

\usepackage[colorlinks=true,urlcolor=darkblue,citecolor=darkblue,linkcolor=darkred,hyperfootnotes=false]{hyperref}

\newcommand{\cc}{\text{c}}
\newcommand{\dd}{\text{d}}
\newcommand{\ee}{\text{e}}

\newcommand{\br}{\text{\bf r}}

\begin{document}

%===================
\title{Activity statistics in a colloidal glass former: experimental evidence for a dynamical transition}

\author{B\'ereng\`ere Abou}
\affiliation{Laboratoire Mati\`ere et Syst\`emes Complexes, UMR 7057 CNRS-P7, Universit\'e Paris Diderot, 10 rue Alice Domon et L\'eonie Duquet, 75205 Paris cedex 13, France}

\author{R\'emy Colin}
\affiliation{Laboratoire Mati\`ere et Syst\`emes Complexes, UMR 7057 CNRS-P7, Universit\'e Paris Diderot, 10 rue Alice Domon et L\'eonie Duquet, 75205 Paris cedex 13, France}

\author{Vivien Lecomte}
\affiliation{LIPhy, Université Grenoble Alpes \& CNRS,  F-38000 Grenoble, France}

\author{Estelle Pitard}
\affiliation{Laboratoire Charles Coulomb, UMR 5221 CNRS-UM2, Universit\'e de Montpellier 2, place Eug\`ene Bataillon, 34095 Montpellier cedex 5, France}

\author{Fr\'ed\'eric van Wijland}
\affiliation{Laboratoire Mati\`ere et Syst\`emes Complexes, UMR 7057 CNRS-P7, Universit\'e Paris Diderot, 10 rue Alice Domon et L\'eonie Duquet, 75205 Paris cedex 13, France}

\affiliation{Department of Chemistry, University of
    California, Berkeley, CA, 94720, USA}

\date{\today}

\begin{abstract}
In a dense colloidal suspension at a volume fraction
   below the glass transition, we follow the
  trajectories of an assembly of tracers over a large time
  window. We define a local activity, which quantifies the local tendency of the system to rearrange. We determine the statistics of the time and space
  integrated activity, and we argue that it develops a low activity tail that comes together with the onset of glassy-like  behavior and heterogeneous dynamics. These rare events may be interpreted as the reflection of an underlying dynamic phase transition.
\end{abstract}


%===================
\maketitle

%===================
\section{Motivations}
In colloidal suspensions, the glass transition refers to the sudden
and sharp increase of viscosity as the volume fraction is increased
above a typical value. In spite of its accepted name, this phenomenon
is not a transition in the standard static sense, according to which a
local ordering spontaneously emerges (like between a liquid and a
crystal), and for which the properties of the microscopic
configurations sampled by the system dramatically change on either
side of the transition. Indeed, when looking at the sampled configurations there
seems not to be deep structural differences between a glass and the
corresponding liquid. In order to characterize and to understand this
phenomenon, a number of tools have been proposed. On the theoretical
side, these include, but are not limited to, intrinsically dynamical
approaches, like the mode-coupling theory, or, more recently, a
phenomenological picture involving the notion of dynamic facilitation (see the review by Ritort and Sollich~\cite{ritort_glassy_2003} or the more recent one by Garrahan, Sollich and Toninelli, chapter 10 in Ref.~[\onlinecite{Berthier:1449083}]). There also exists a purely statics-based proposal,
namely the random first order theory (RFOT) scenario~\cite{PhysRevA.40.1045}, which
depicts the glassy state as originating from a genuinely thermodynamic
phenomenon. This theoretical approach is backed by direct equilibrium
statistical-mechanical calculations on microscopic systems of
interacting particles (either molecular or colloidal), unlike the
dynamical facilitation picture which relies only on kinetic
rules. Experimental works aiming at sorting out the appropriate
theoretical picture, by testing model assumptions and predictions, are
scarce. A remarkable work by Gokhale {\it et
  al.}~\cite{gokhale_growing_2014,PhysRevLett.116.068305} does exactly this, by focusing on
the nature of local excitations in space and time and by probing
point-to-set correlations, and showing their consistency with the
dynamical facilitation picture.
%A trademark of dynamical facilitation is the existence of dynamical heterogeneities, which are independent regions of space where coherent correlated motion occurs and controls the system's relaxation. These dynamical heterogeneities have been the focus of intense theoretical and experimental 
%scrutiny and they have been shown to account for the time-behavior of nonlinear susceptibilities~\cite{berthier_direct_2005}.
 Dynamical facilitation has recently been
 shown~\cite{garrahan_dynamical_2007, hedges_dynamic_2009,
   pitard_dynamic_2011} to go hand in hand with an unusual signature
 behavior of the distribution of a number of macroscopic quantities
 built from the local observables considered in Ref.~[\onlinecite{gokhale_growing_2014}]. The twist in defining these macroscopic
 observables -- generically christened {\it activity} -- is to build
 them by not only summing local spatial quantities over the whole
 system, but above all to integrate the latter over the course of a
 large time interval, so that these exhibit both space and time
 extensivity. Dealing with the fluctuation properties of space and
 time extensive observables is the business of the so-called thermodynamic formalism~\cite{ruelle} based
 on spatio-temporal histories. Here, our purpose is neither to go deeper
 into the mathematical formalism of dynamic facilitation, nor to address a direct local quantification of space and time correlations between rearrangement events. Instead we show these ideas can be practically implemented. It is not only
 experimentally feasible to measure such a distribution of a space-time extensive observable in a system of
 interacting colloids but also that telling information can be
 gathered from such measurements. Concomitantly to the present work, Pinchaipat {\it et al.}~\cite{2016arXiv160900327P}, with poly-methylmethacrylate (PMMA) colloidal particles, also found interesting features in the study of time-extensive physical observables.

Our experimental system is a dense suspension of thermosensitive
microgels, in which a low density of tracer latex beads has been
uniformly dispersed. We track the motion of the tracers
in space and time, thus gathering a set of full trajectories
$\br_j(t)$ corresponding to each individual tracer $j$. These are the
primary material of our study. These tracers have been
shown~\cite{colin_spatially_2011} to be accurate probes for dynamical
heterogeneities. The duration $t_\text{obs}$ of a trajectory is sliced
into $M$ lapses of duration $\Delta t$, which we choose to be of the
order of the time it takes a fluctuation to drive a tracer away by a
fraction of its diameter. This is the instanton time introduced in Ref.~[\onlinecite{hedges_dynamic_2009}], further discussed in~[\onlinecite{keys_excitations_2011}] and concretely used in~[\onlinecite{gokhale_growing_2014}] to analyze a binary mixture of silica
colloids. Following Speck
and Chandler~\cite{speck_constrained_2012} we define a space and time
{\it activity} $K$ as a functional of
the trajectories of the observed tracer $i$
\begin{equation}\label{defactivity}
K[\br_i](t_{obs})=\sum_{j=1}^{M=t_\text{obs}/\Delta t}\!\!\!\!\Theta\left(||\br_i(j\Delta t)-\br_i((j-1)\Delta t)||-a\right)
\end{equation}
where $\Theta(x)$ is the step function and $a$ is a length scale
of the order of a fraction of the particle diameter. The purpose of
this step function activity $K$ is to count the number of events in which
a tracer has been able, in a time $\Delta t$, to hop away from its
local environment by a distance $a$. Being the sum of a large number
$M$ of local random events (albeit correlated) we expect that the
relative fluctuations in $K$ will drop as $\frac{1}{\sqrt{M}}$. But of
course $M$ has to be large enough so that $K$ captures the space and
time-correlated motion of the tracers. The latter motion will be mediated by the spatially and temporally heterogeneous dynamics. Recent discussions connecting the areas of the system that witness cooperative rearrangement, the structure of facilitated excitations, or soft spots, can be found in Ref.~[\onlinecite{keys_calorimetric_2013}] for instance. Our purpose in
this work is to present the distribution of the activity $K$ as the
experiment (where we observe a tracer over time $t_\text{obs}=M\Delta
t$) is repeated a large number of times and over all available
tracers. Beyond typical (Gaussian, central-limit related)
fluctuations, we aim at quantifying deviations from the Gaussian and
to show that these are consistent with the theoretical predictions of Ref.~[\onlinecite{hedges_dynamic_2009}]. Namely, the distribution of a global,
space and time integrated, physical observable, has features that
reflect the space-wise and time-wise local peculiarities of the glassy
state.

In the following we describe our experimental system and imaging tool. Then we provide a theoretical section delving into the details of why rare events tail in the activity distribution is of interest. Our experimental results are then presented and analyzed.

\section{Materials and Methods}

Our model glass consists of a suspension of thermosensitive microgels,
made of the amphiphilic polymer poly(N-isopropylacrylamide)
(pNIPAm). The particle’s radii can be reversibly tuned by changing the
temperature of the suspension~\cite{Questioning2015}. When the temperature decreases, the particle radius increases, and so does the volume fraction of the
suspension. This enables to explore the various states of the suspension from the liquid to the supercooled liquid state and the glassy state~\cite{colin_spatially_2011}. This is described in Appendix~\ref{app:pNipam}.
In our study, we focused on
dense suspensions at two temperatures, $T=29^\circ$C and
$T=27^\circ$C, corresponding to two effective volume
fractions\cite{Questioning2015} -- namely, $\phi_{29} =0.44 \,(\pm 0.02)$ and
$\phi_{27} =0.54 \,(\pm 0.02)$ -- below the glass transition for these
soft colloids ($\phi_g$ between $0.65$ and $0.70$~\cite{colin_suspensions_2012}). The diameters of the particles are
$\sigma_{29}=0.887\pm0.005 \,\mu$m at $T= 29^\circ$C and
$\sigma_{27}=0.945 \pm 0.010 \,\mu$m at $T= 27^\circ$C. 
The two suspensions are in the supercooled liquid state, as can be demonstrated in 
Figure~\ref{msd} and in Appendix~\ref{app:pNipam} which describes 
the phase diagram of the system. The relative relaxation times of the two suspensions are 
respectively ${\tau_r}_{29}=128\pm7$ and ${\tau_r}_{27}=649\pm 32$, obviously between the liquid state 
(where PDFs of displacement are Gaussian) and the glass state (where there is aging). 
 The $T= 27^\circ$C is in a deeper supercooled state than the $T= 29^\circ$C suspension.
 %The diameters of the particles are
%$\sigma_{29}=0.887\pm0.005 \,\mu$m at $T= 29^\circ$C and
%$\sigma_{27}=0.945 \pm 0.010 \,\mu$m at $T= 27^\circ$C. 
%
The
suspensions were seeded with a low fraction ($0.1\%$) of polystyrene
beads ($0.994 \,\mu$m in diameter) which serve as tracers of the
dynamics. The suspension is injected into a $3 \times 3$ mm$^2$
chamber made of a microscope slide an a coverslip separated by a $250
\,\mu$m thick adhesive spacer. The chamber was sealed with araldite
glue to avoid evaporation and contamination. The samples are observed under bright 
field transmitted light microscopy at $100X$
magnification and recorded using a Eosens CMOS camera (field of view
$512 \times 512$ px$^2$, $1$ px corresponds to $0.138 \,\mu$m). The temperature of the
suspension was maintained constant using a Bioptechs objective heater
acting on the sample through the immersion oil. 
Because of the difference in the
diffusion rate when varying volume fraction, images were collected every $0.2$ s during $160$ s for
the $T= 29^\circ$C suspension ($800$ images per movie) and every $0.5$ s during
$500$ s at $T= 27^\circ$C ($1000$ images per movie). For each volume fraction
 (temperature $T= 29^\circ$C and $T= 27^\circ$C), $10$ independent movies were acquired. The fraction 
 of tracers added to the soft particles suspension provides 
between $50$ and $100$ tracers in the field of view, depending on the movie. The
region of observation was chosen at least $100 \,\mu$m away from the
sample edges to avoid boundary effects. 
A self-written analysis software allowed us to track the tracer
positions $x(t)$ ans $y(t)$, close to the focus plane, and to
calculate all the quantities presented in the following: mean-squared
displacement, activity, variance and skewness. For each probe $j$ in the field of view, 
the time-averaged quantity was calculated. For each movie, we ensemble-averaged the considered quantity 
over all the probes present in the field of view. The quantity was finally averaged over the $10$ 
independent movies available 
for each volume fraction. This allowed us to accumulate a 
large statistical ensemble for each set of data.

\section{Dynamical activity}
\subsection{Activity of a diffusive ideal gas}\label{ideal}

As a preliminary investigation, and to set up a reference system, it
is instructive to formulate and answer the questions we are interested
in for a Brownian particle. We refer the reader to the
Appendix~\ref{app:A_BM} for mathematical details. Our purely diffusive
Brownian particle has diffusion coefficient $D$. We ask how the
activity of this particle with trajectory $\br(t)$ is distributed in
$d=2$ and $d=3$ dimensions.

The average activity $\langle K\rangle/M$ of this particle over the
time histories reads, in terms of the dimensionless scaling variable
$u=\frac{a}{\sqrt{D\Delta t}}$:
\begin{equation}\label{meanact}
\frac{\langle K\rangle}{M}=\left\{\begin{array}{ll}
\text{erfc}\left(\frac{u}{2}\right)+\frac{\ee^{-\frac{u^2}{4}} u}{\sqrt{\pi }}&{(d=3)}\\
\ee^{-\frac{u^2}{4}}&{(d=2)}
\end{array}\right.
\end{equation}
with the right hand side in~\eqref{meanact} bounded by 0 and 1. We
find it of interest to focus on the
normalized third cumulant, otherwise known as the skewness $\kappa_3$ of the
distribution, to quantify the first nontrivial signature of a
deviation with respect to the Gaussian distribution.
%the first nontrivial signature of a
%deviation with respect to the Gaussian distribution, namely the
%normalized third cumulant, otherwise known as the skewness of the
%distribution.
The skewness $\kappa_3$, which is a measure of cubic
correlations, reads in dimension $3$:
\begin{small}%
  \begin{equation}
  \hspace*{-12mm}
  \kappa_3=\frac{1}{\sqrt{M}}\frac{\ee^{-\frac{u^2}{4}}
    \left(\sqrt{\pi } \ee^{\frac{u^2}{4}} \left(2
    \text{erf}\left(\frac{u}{2}\right)-1\right)-2
    u\right)}{\sqrt{\sqrt{\pi } \ee^{-\frac{u^2}{4}} u \left(2
      \text{erf}\left(\frac{u}{2}\right)-1\right)-\pi
      \left(\text{erf}\left(\frac{u}{2}\right)-1\right)
      \text{erf}\left(\frac{u}{2}\right)-\ee^{-\frac{u^2}{2}} u^2}}
\label{eq:skewnessd3}
\end{equation}%
\end{small}%
A similar calculation carried out in space dimension 2 gives:
\begin{equation}
\kappa_3=\frac{1}{\sqrt{M}}\frac{-2+\ee^{\frac{u^2}{4}}}{\sqrt{-1+\ee^{\frac{u^2}{4}}}}
\label{eq:skewnessd2}
\end{equation}
We will use these expressions for the skewness in Section~\ref{meat}
as a benchmark to assess the deviation from the diffusive ideal gas behavior.

The idea of examining cubic correlations has emerged recently as a
useful tool for the analysis of glassy systems. The work carried out
by Crauste-Thibierge {\it et al.}~\cite{PhysRevLett.104.165703} does
exactly that as a function of time (in a similar spirit to
studies of the nonlinear susceptibility $\chi_4(t)$). The observable
$\kappa_3$ captures instead time-integrated
aspects.

\subsection{What are the expected changes in the presence of interactions ?}

The definition of the activity in our dense suspension with soft
repulsive interactions is exactly identical to that appearing in the
previous subsection~\ref{ideal}. However the parameters $a$ and
$\Delta t$ entering the activity cannot form a single variable based
on the diffusive scaling. The interacting system carries its own space
and time scales. In such dense systems, the reference spatial scale is the
range of the interaction potential (or the ``size'' of the particles);
However, a hierarchy of relevant time-scales show up. The shortest one
tells us about local vibrations of the particles around their
equilibrium positions. Fast-rattling about a local equilibrium
position is not what we are interested in. Instead, the intermediate
time-scale of interest to us is related to the time required for a
particle to participate in a cooperative rearrangement at the particle
scale, leading to another local equilibrium position.
%Such areas of the system that witness cooperative rearrangement are often called excitations or soft spots in the language of dynamic facilitation~\cite{keys_calorimetric_2013}. 
Rearrangement events should decrease in frequency and density as the system approaches the glass transition. This is the reason why the macroscopic relaxation rate is seen to increase as temperature is lowered.

While the microscopic mechanism behind the emergence of independent
localized excitation patches is unknown, a vast body of theoretical
work has been devoted to the study of model systems of excitations
with facilitated dynamics~\cite{ritort_glassy_2003}. Soft spots are believed to
display equilibrium correlations between themselves. However, they
exhibit unusually correlated dynamics. One way to capture this
property is to investigate the histories of the system configurations
over a large time interval. In practice one way to do this is to
consider a time extensive quantity and to investigate its
fluctuations. The activity is indeed a most relevant measure of the
integrated number of excitations over space and time.
%yet this has been quantified with various microscopic definitions. 

In the theoretical literature, the space and time integrated number of
excitations up until some observation time $t_\text{obs}$ is the
quantity used~\cite{garrahan_dynamical_2007} in the study of systems
with kinetic constraints expressing dynamical facilitation. The
activity introduced by Hedges {\it al.}~\cite{hedges_dynamic_2009}
which we described earlier in Eq.~\eqref{defactivity} is a proposal
adapted to realistic molecular systems. Both in the model lattice
systems and in the realistic molecular systems, the probability
distribution of the activity displays a double peak structure that is
a trademark of supercooled
liquids\,\footnote{Note that in a simple fluid phase, a
    single peak is observed.}.

In standard equilibrium statistical mechanics, a double peak for the
  distribution of an order parameter signals a first order
transition between two coexisting phases of a system. Given the dynamical nature of the activity, that its
distribution displays a double peak signals the coexistence of two dynamical time evolutions of the
  system. Hence the ``dynamical first order transition''
terminology that theoretical works have adopted.

The prediction~\cite{garrahan_dynamical_2007,
  hedges_dynamic_2009,pitard_dynamic_2011} that we wish to investigate here 
at the experimental level is that the distribution of the activity is a signature of an
underlying transition between two dynamical phases: an
equilibrium-like phase with homogeneous dynamics, and an
ergodicity-breaking phase with slow dynamics and
  low activity.
We emphasis that the transition that we study is intrinsically dynamical in the sense that, in contrast to previous experimental works~\cite{chikkadi_shear_2014}, it does not correspond to an underlying standard phase transition -- such transitions being indeed known~\cite{lecomte_chaotic_2005,lecomte_thermodynamic_2007} to induce dynamical transitions in a generic manner.

However, the specifics of our experimental analysis differ from the numerical protocol adopted in
previous theoretical works. We have decided to monitor tracers
individually and to define a tracer-dependent activity, as is explicit
in Eq.~\eqref{defactivity}. At the level of individual tracers a dynamical first order
transition is suggested by the emergence of a secondary peak
around the low-activity phase. This can be viewed as a precursor
of the real $N$-body effect that would appear if we monitored
collectively the set of tracers. The existence of a
low-activity dynamical phase will be manifest in the emergence of a
negative excess in the skewness of the activity distribution. Indeed,
a negative skewness means a fatter tail for atypically small events.

\section{Results}\label{meat}

\begin{figure}[t]
\centering
\includegraphics[width=.7\columnwidth]{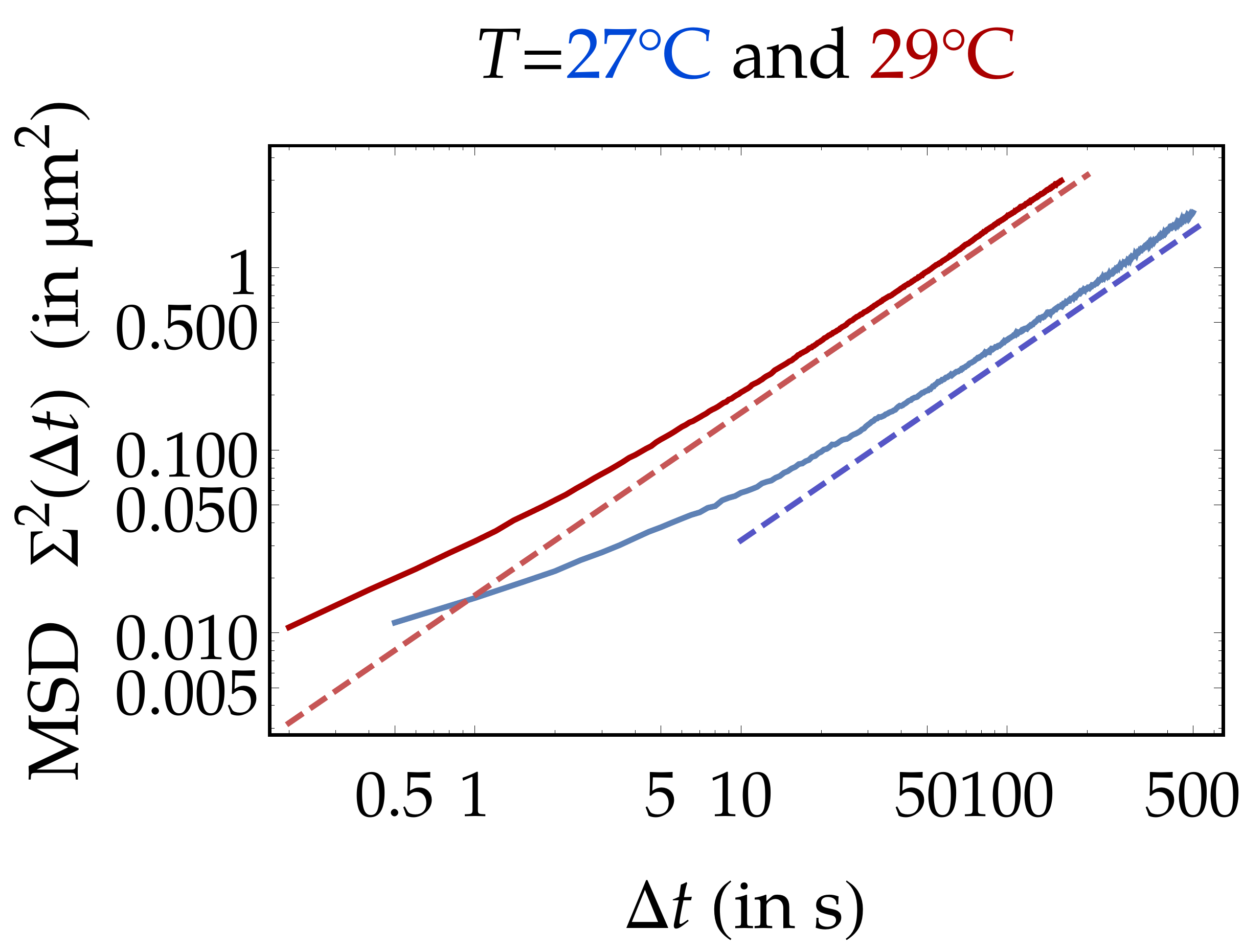}
\caption{Mean-squared displacement $\Sigma^2\!(\Delta
  t)=\langle(\br_j(t+\Delta t)-\br_j(t) ))^2\rangle_{t,j}$ of the
  tracers in the dense microgel suspension at temperatures
  $T=27^\circ\text{C}$ (blue) and $T=29^\circ\text{C}$ (red), as a function of the
  lag time $\Delta t$. For both temperatures, we observe a long-time diffusive behavior 
  (alpha relaxation), preceded by a sub-diffusive behavior 
  at intermediate time-scales. At $T=27^\circ\text{C}$, 
the suspension dynamics is much slower than at $T=29^\circ\text{C}$.
The short-time diffusive behavior 
is not presented here, and can be seen in Ref.~[\onlinecite{colin_suspensions_2012}]. 
In the diffusive regime, the MSD slope is 0.017~$\mu$m$^2$/s for the $T=29^\circ\text{C}$ data and 0.0032~$\mu$m$^2$/s for the $T=27^\circ\text{C}$ data.
The relative relaxation times are respectively 
${\tau_r}_{29}=128\pm7$ and ${\tau_r}_{27}=649\pm 32$, which can be 
compared to the curves of the pNipam phase diagram of 
Appendix~\ref{app:pNipam}. Both suspensions are in the supercooled liquid phase. 
\label{msd}}
\end{figure}
%%
%% from /home/lecomtev/recherche/experiments_berangere-remy/analyse-resultats-remy.nb
%% from /home/lecomtev/recherche/experiments_berangere-remy/analyse-resultats-remy_second-round.nb
%%
\begin{figure}[t]
\centering
\includegraphics[width=.49\columnwidth]{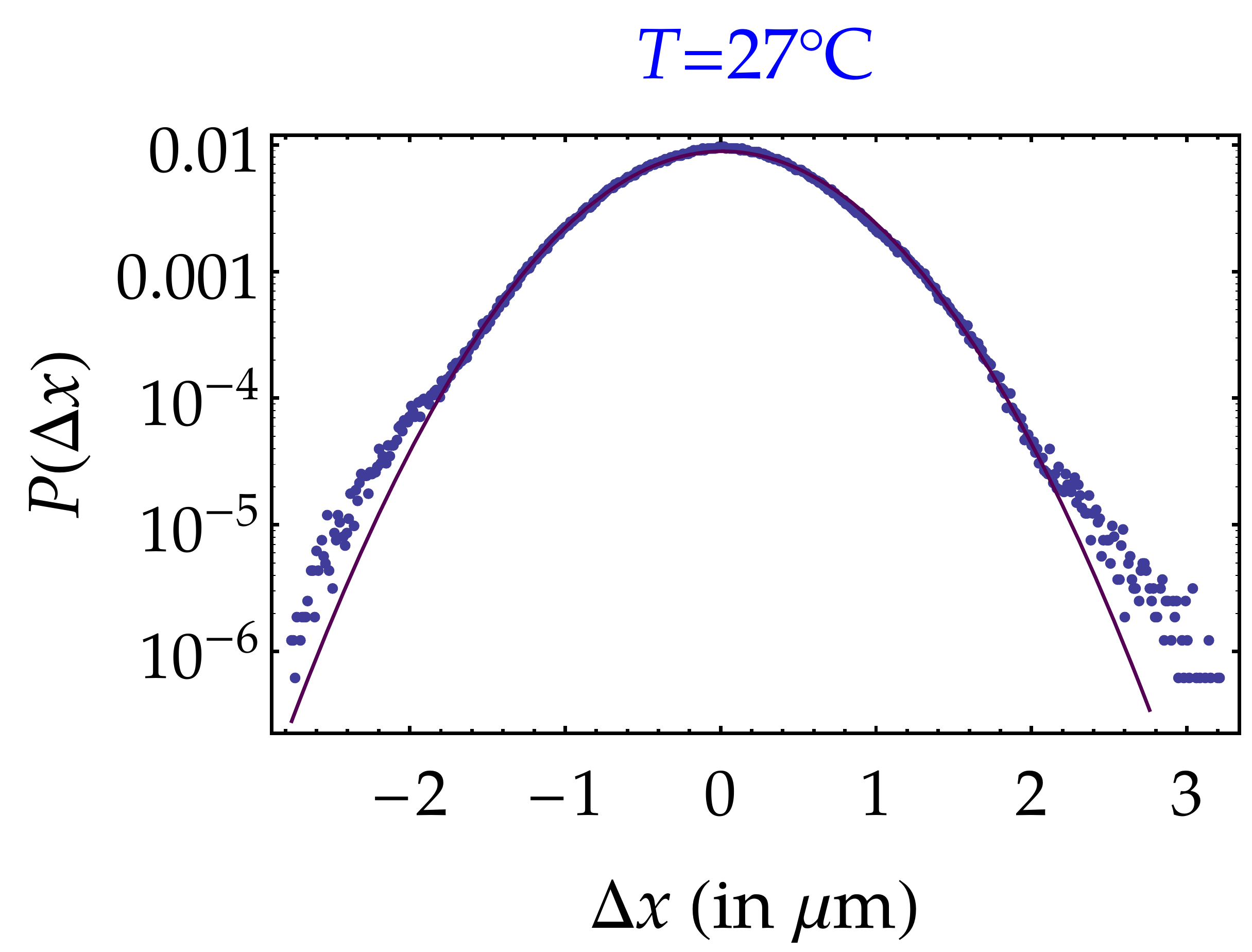}
\includegraphics[width=.49\columnwidth]{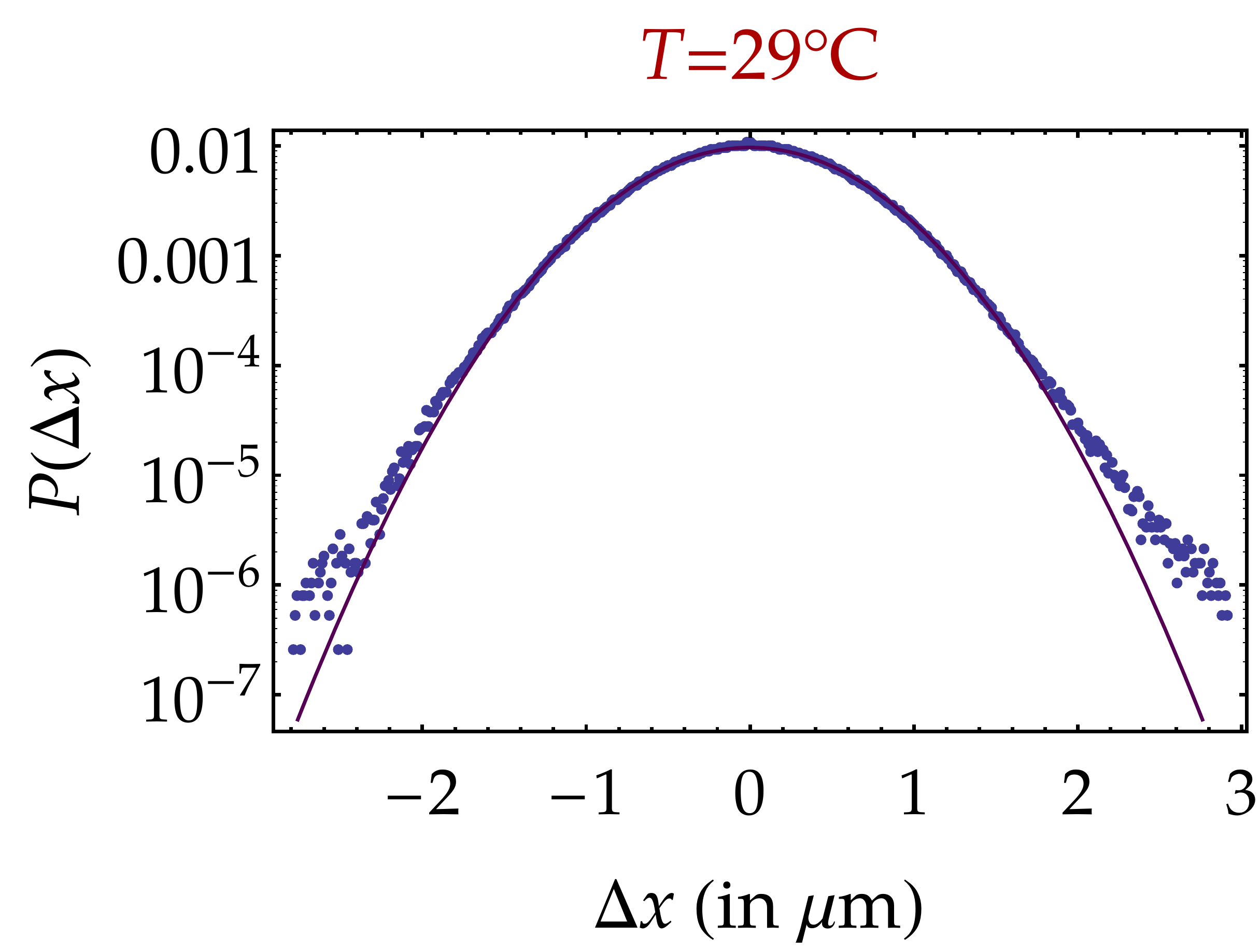}
\caption{Probability distribution function of the displacement $\Delta
  x$ of the tracers in the dense microgel suspension during $\Delta
  t_{\sigma}$, at temperatures $T=27^\circ\text{C}$ (left) and
  $T=29^\circ\text{C}$ (right). For each temperature, the time lapse $\Delta
  t_{\sigma}$ is fixed by imposing $\Sigma\!(\Delta
  t_{\sigma})=\frac 12\sigma$, with $\sigma$ the microgel diameter. 
  One finds $\Delta t_{\sigma}\simeq 53\:\text{s}$ and $\Delta t_{\sigma}\simeq 9.4\:\text{s}$ for $T=27^\circ\text{C}$ and $T=29^\circ\text{C}$ respectively.
  The PDFs
  departs from a Gaussian distribution for both temperatures, as
  extensively discussed in Ref.~[\onlinecite{colin_spatially_2011}]. The solid line represents the Gaussian distribution. 
%
  %GIVE HERE THE  EXACT VALUE IN SECONDS (point de repere par rapport aux autres Delta  t utilises)
% [NOT ANYMORE] The horizontal axis is in pixels, defined by $1 {\rm pixel}= 138 {\rm nm}$.
\label{pdfDeltax}}
\end{figure}
%%
%% from /home/lecomtev/recherche/experiments_berangere-remy/analyse-resultats-remy.nb
%% from /home/lecomtev/recherche/experiments_berangere-remy/analyse-resultats-remy_second-round.nb
%%

\subsection{Mean-squared displacement, Average activity}

Fig.~\ref{msd} shows the mean-squared displacement (MSD) 
\begin{equation}
\Sigma^2\!(\Delta t)=\big\langle(\br_j(t+\Delta t)-\br_j (t))^2\big\rangle_{t,j}
\label{eq:defMSD}
\end{equation}
of the tracers as a function of the lag time $\Delta t$, in the dense
suspension at two temperatures (thus combining the data from both
space directions $x$ and $y$). The average runs over the tracers $j$
and the reference time~$t$. We use Fig.~\ref{msd} to extract the value
$\Delta t_{\sigma}$, such that the mean-squared displacement is equal
to the squared diameter of a microgel, $\Sigma^2\!(\Delta
t_{\sigma})=\sigma^2$. This value $\Delta t_{\sigma}$ is used in the
following for the computation of the probability distribution function (PDF) of the displacement.
%
%We emphasise that our interest goes to the long-time behaviour (alpha relaxation of the MSD) since it is the relevant time scale for our purpose. 
%
%For the same experimental configuration, the beta relaxation (short-time diffusive behaviour) can be 

%observed at shorter time scales~\cite{colin_suspensions_2012}.
%
%The MSD does not allow one to clearly characterise the (possible) overcooled (?) nature of the system, and our aim is to devise better observables to identify such a state.

%Correction for an experimental mechanical drift is implemented.
%These data are to be compared with the diameter of the colloids at
%these temperatures, namely $\sigma(27)=1.88\;\mu\text{m}$ and
%$\sigma(29)=1.79\;\mu\text{m}$.

%\textcolor{red}{It matters whether we talk about the 1d (projected) or 2d MSD.}
%checked: it is the 2D one, corresponding to the formula $\langle(\br(t+\Delta t)-\br(t)^2\rangle$

The PDFs of the displacement shown in
Fig.~\ref{pdfDeltax}, deviates from a Gaussian distribution for each temperature studied. 
This non-Gaussian 
behavior of the PDF was extensively discussed 
in Ref.~[\onlinecite{colin_spatially_2011}] in terms of local
heterogeneities of the diffusion coefficient, and was shown to be the signature 
of a supercooled regime. 
Deviations from the Gaussian
emerge after $\Delta x$ exceeds a threshold value $a$ of the order of
$2\sigma$, which corresponds to atypical
events. Here in contrast, we wish to characterize the emergence of dynamical
heterogeneities, at scales smaller that
$2\sigma$ and for non-static observables (intermediate scales).

%In contrast, we wish to characterize, at scales smaller that $2\sigma$ and for
%non-static observables, the emergence of dynamical heterogeneities as the system becomes more glassy.

\begin{figure}[t]
\centering
\includegraphics[width=.75\columnwidth]{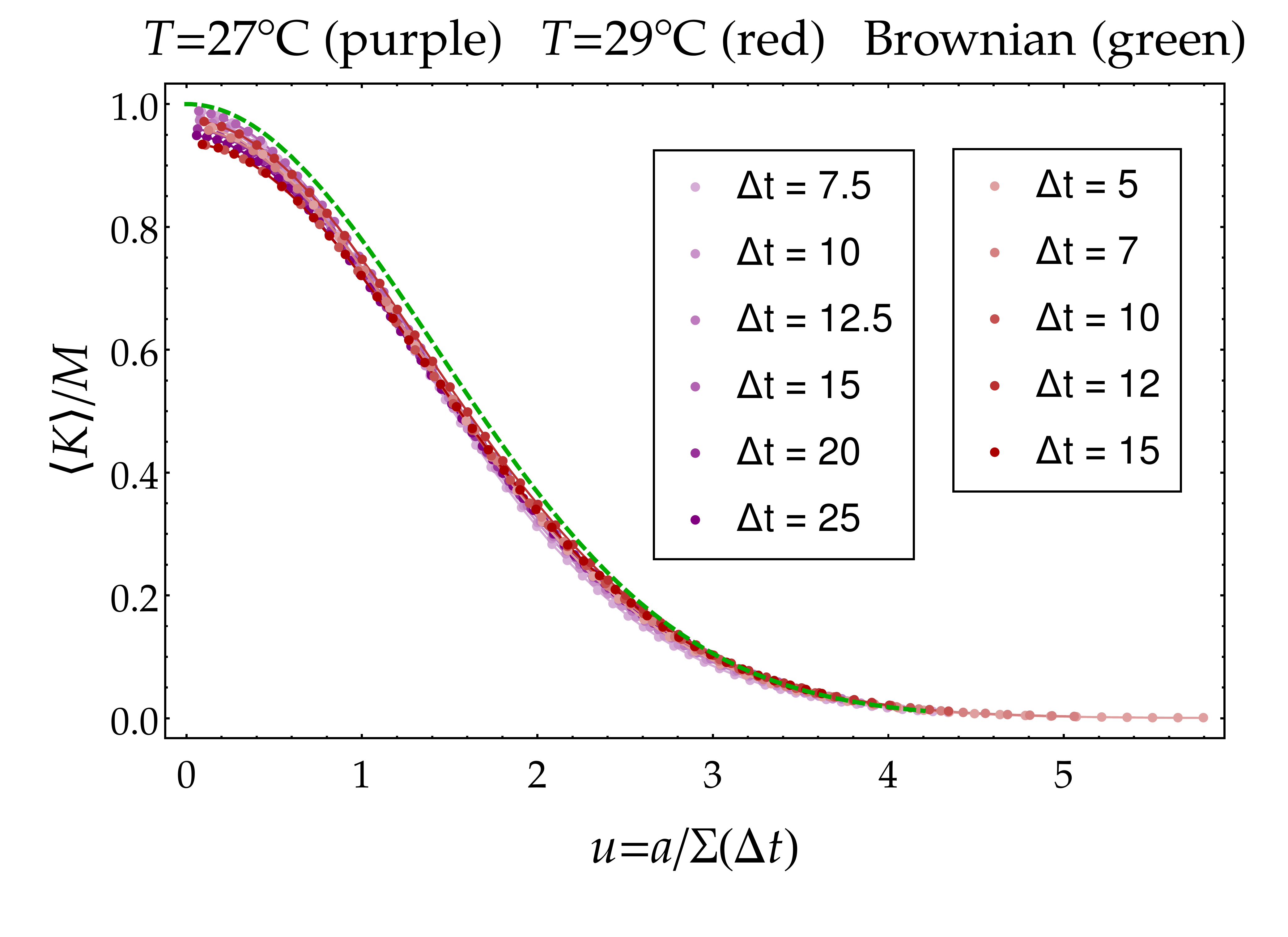}
\caption{Average activity $\langle K\rangle/M$ as a function of $u={a}/\Sigma(\Delta t)$ at temperatures $T=29^\circ\text{C}$ (red) and $T=27^\circ\text{C}$ (purple). The dashed green curve stands for the Brownian counterpart~\eqref{meanact} in dimension $d=2$. For each temperature/color, the various curves correspond to different choices of $\Delta t$. At $T=27^\circ\text{C}$ the $\Delta t$'s are such that $\Sigma(\Delta t)/\sigma$ lies between $0.09$ and $0.16$; and between $0.17$ and $0.29$ at $T=29^\circ\text{C}$.
   \label{activityRemy}
}
\end{figure}
%%
%% from /home/lecomtev/recherche/experiments_berangere-remy/donnees-remy_analysis-december-2014/analyse-resultats-remy_activite-statistique-T27_dec2014.nb (joint graph)
%% from /home/lecomtev/recherche/experiments_berangere-remy/donnees-remy_analysis-december-2014/analyse-resultats-remy_activite-statistique-T29_dec2014.nb
%%

The average activity $\langle K\rangle/M$ of the tracers immersed in
the dense suspension at two temperatures, is shown in
Fig.~\ref{activityRemy} as a function of $u=a/\Sigma(\Delta t)$, for
different values of $\Delta t$. The average is computed over all the
tracers present in the sample, as it is the most direct way
experimentally to calculate averages over time histories.
One remarks that the experimental results are close to the average
activity of a purely diffusive Brownian particle, given
by~\eqref{meanact}, in dimension $d=2$\footnote{The effective dimension of the tracer dynamics is $d=2$ and not $d=3$ since the tracking procedure selects trajectories which remain in the focus plane during the experimental duration of the measurements.}.
One infers from Fig.~\ref{activityRemy} that the average activity is a
poor observable to discriminate between dynamical regimes: Although they exhibit different dynamics (probed by the PDFs for example), both temperature suspensions and the purely diffusive Brownian particle display a similar activity.
Hence, in the next section, we will search for a macroscopic evidence
of dynamical heterogeneities by looking at the full distribution of
the activity.

%\textcolor{red}{We would like to see the variances of $P(K)$ normalized before turning to the skewness.}
\subsection{Histograms  of the activity}
\label{ssec:histogramsK}

\begin{figure}[h]
\centering
\includegraphics[width=.32\columnwidth]{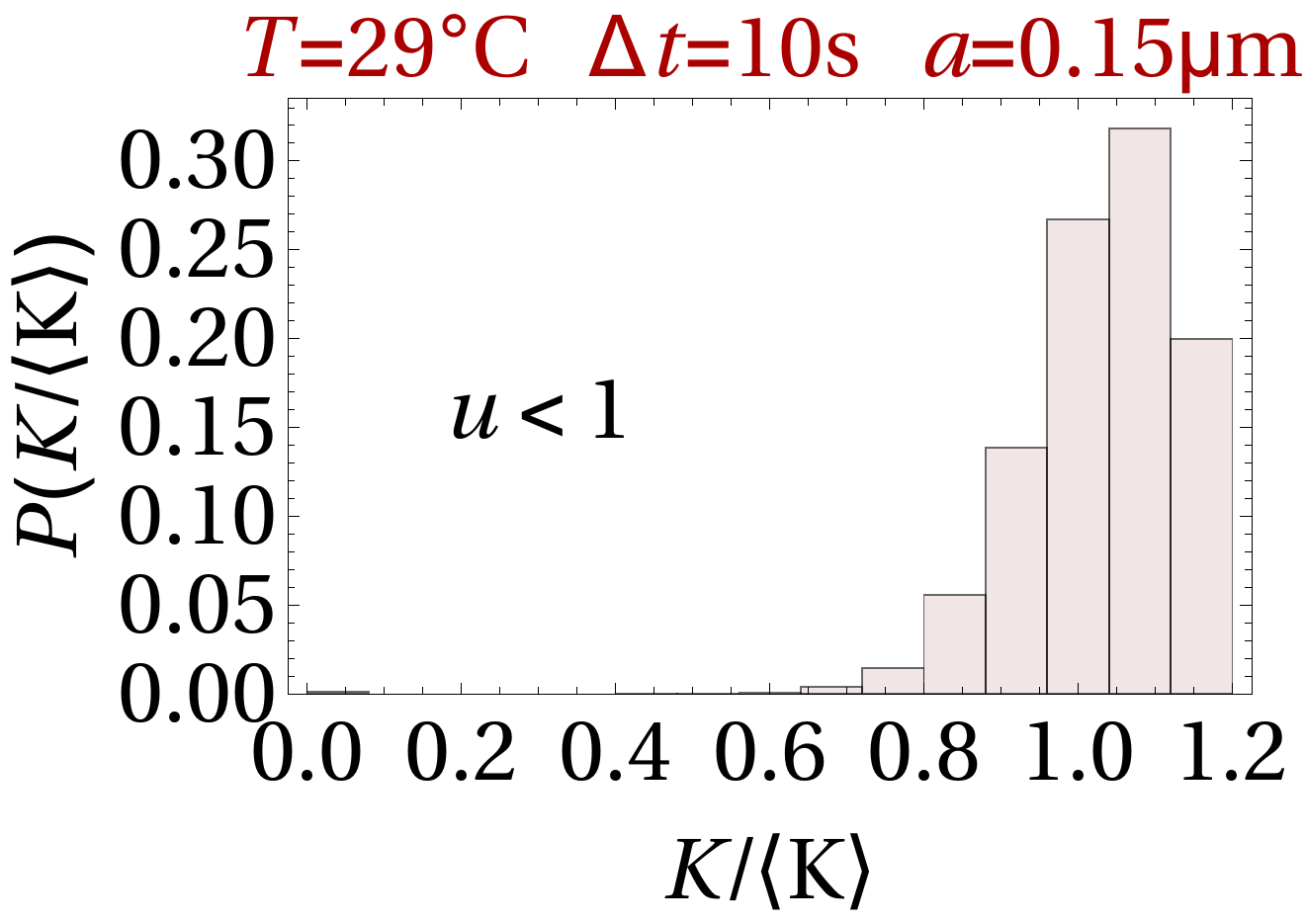}
\includegraphics[width=.32\columnwidth]{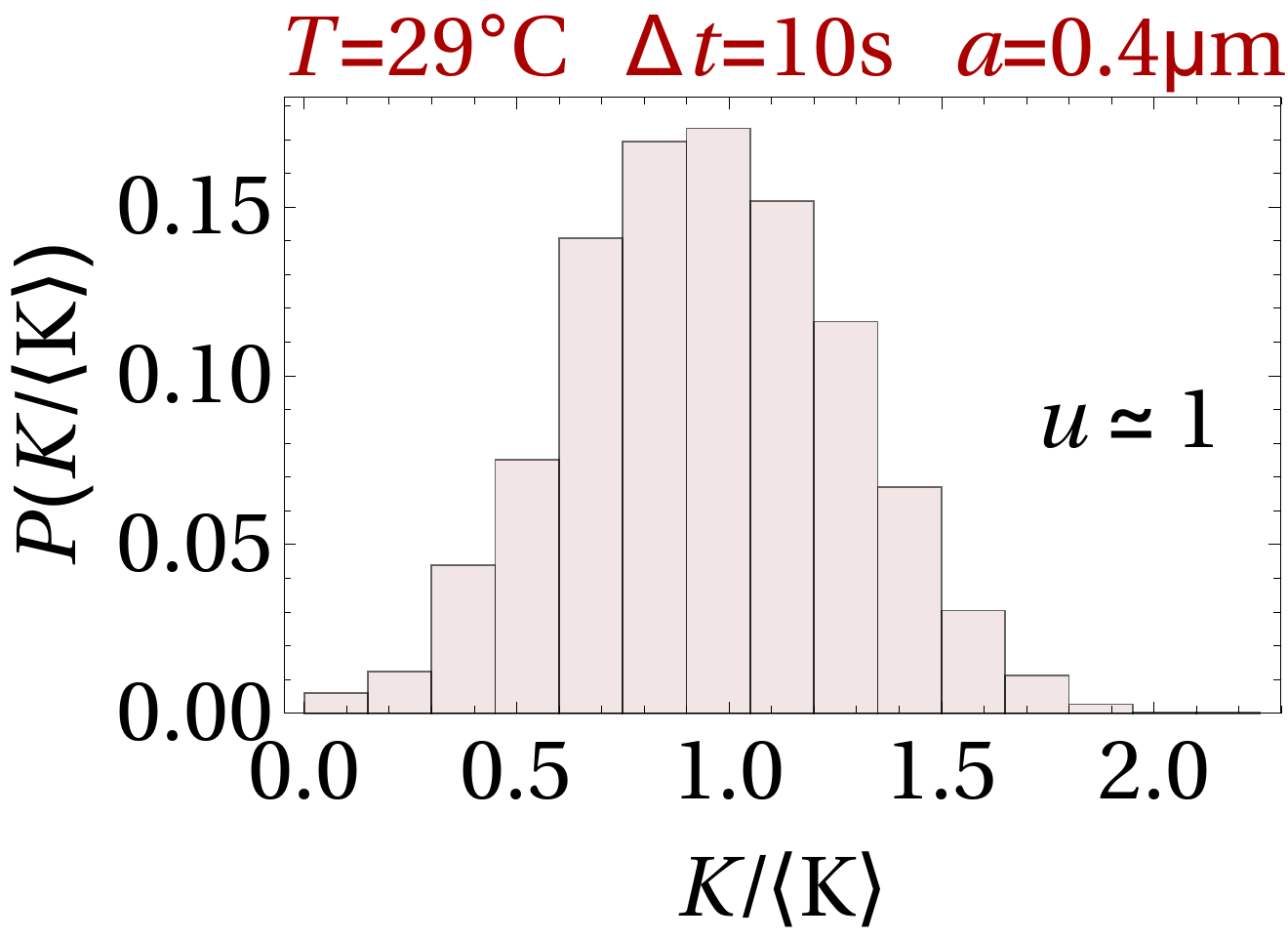}
\includegraphics[width=.32\columnwidth]{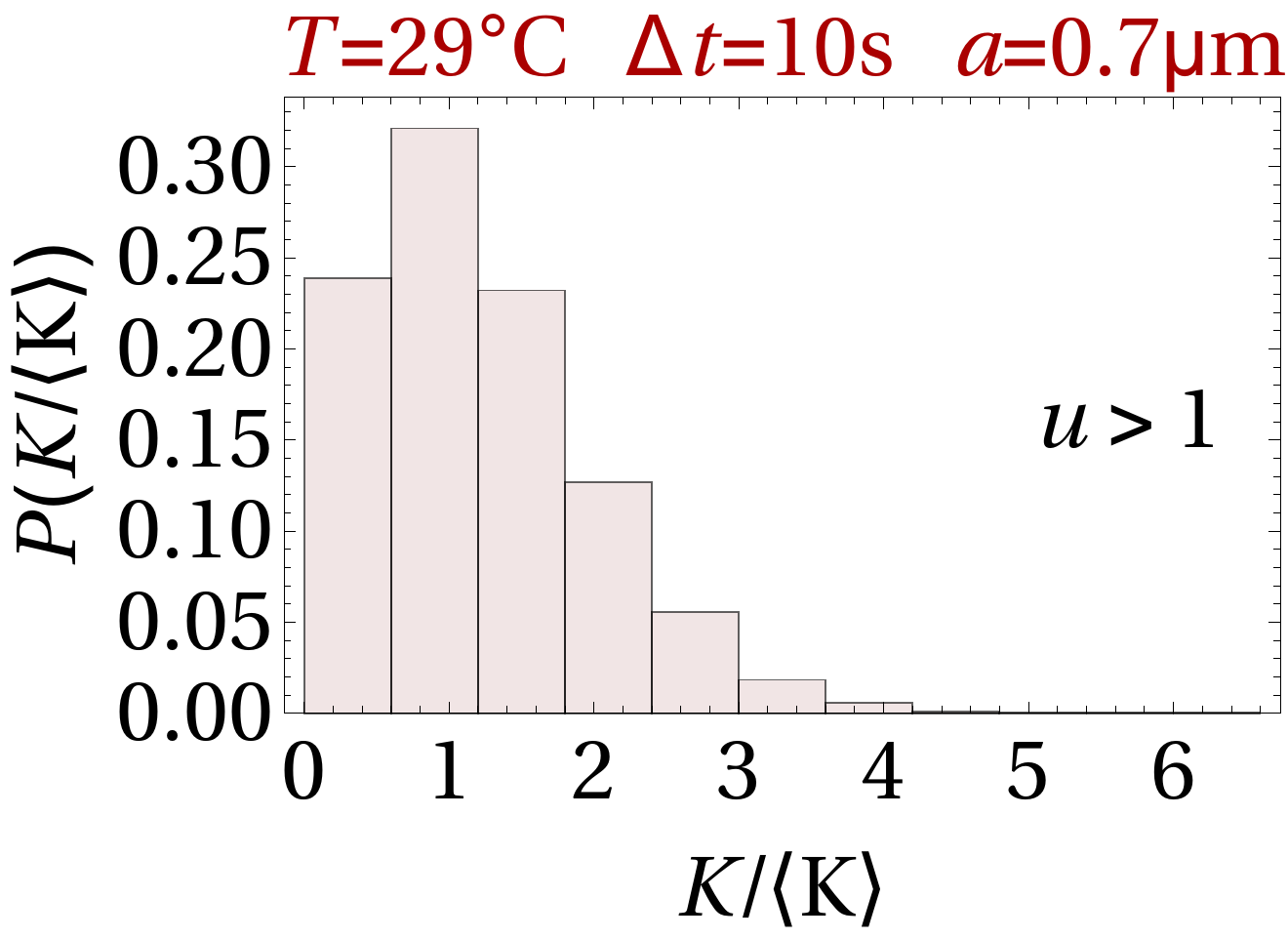}
\caption{From left to right, histogram of the activity at $T=29^\circ\text{C}$ normalized by its average for $a=0.084\sigma$, $a=0.22\sigma$ and $a=0.39\sigma$, respectively.
The corresponding values of $u$ are then respectively: 0.32, 0.87 and 1.52, and those of $\langle K \rangle$: 13.4, 6.83 and 1.52\,.
The time lapse $\Delta t$ is $10\,\text{s}$. The more symmetric histogram (center) is at value of $a$ of the same order as $\Sigma(\Delta t)$. 
%PEUT ON METTRE sur chaque graphe, $u<1$, $u>1$ et $u\simeq 1$.
   \label{hist29}}
\end{figure}

\begin{figure}[h]
\centering
\includegraphics[height=.23\columnwidth]{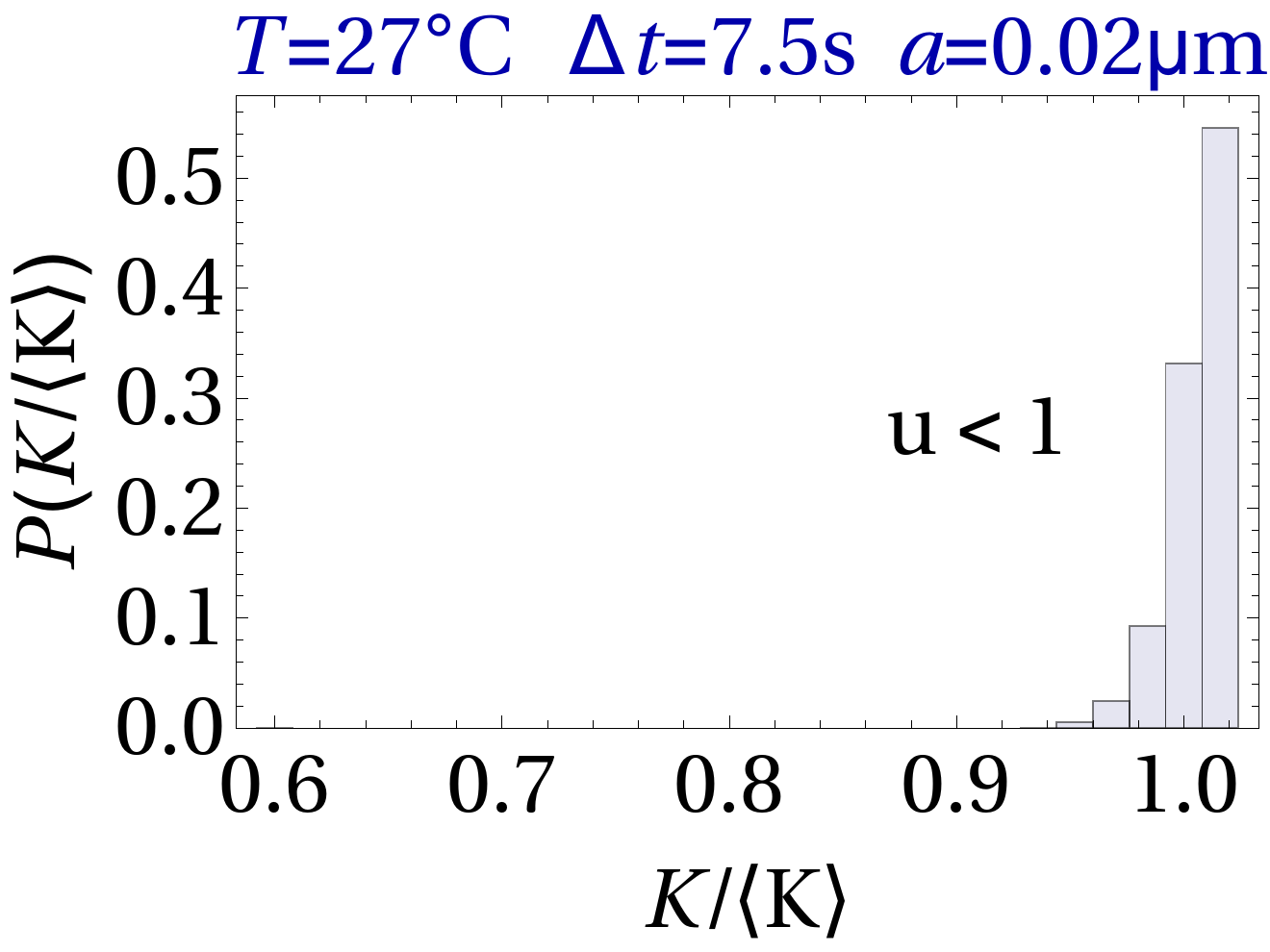}\hfill
\includegraphics[height=.23\columnwidth]{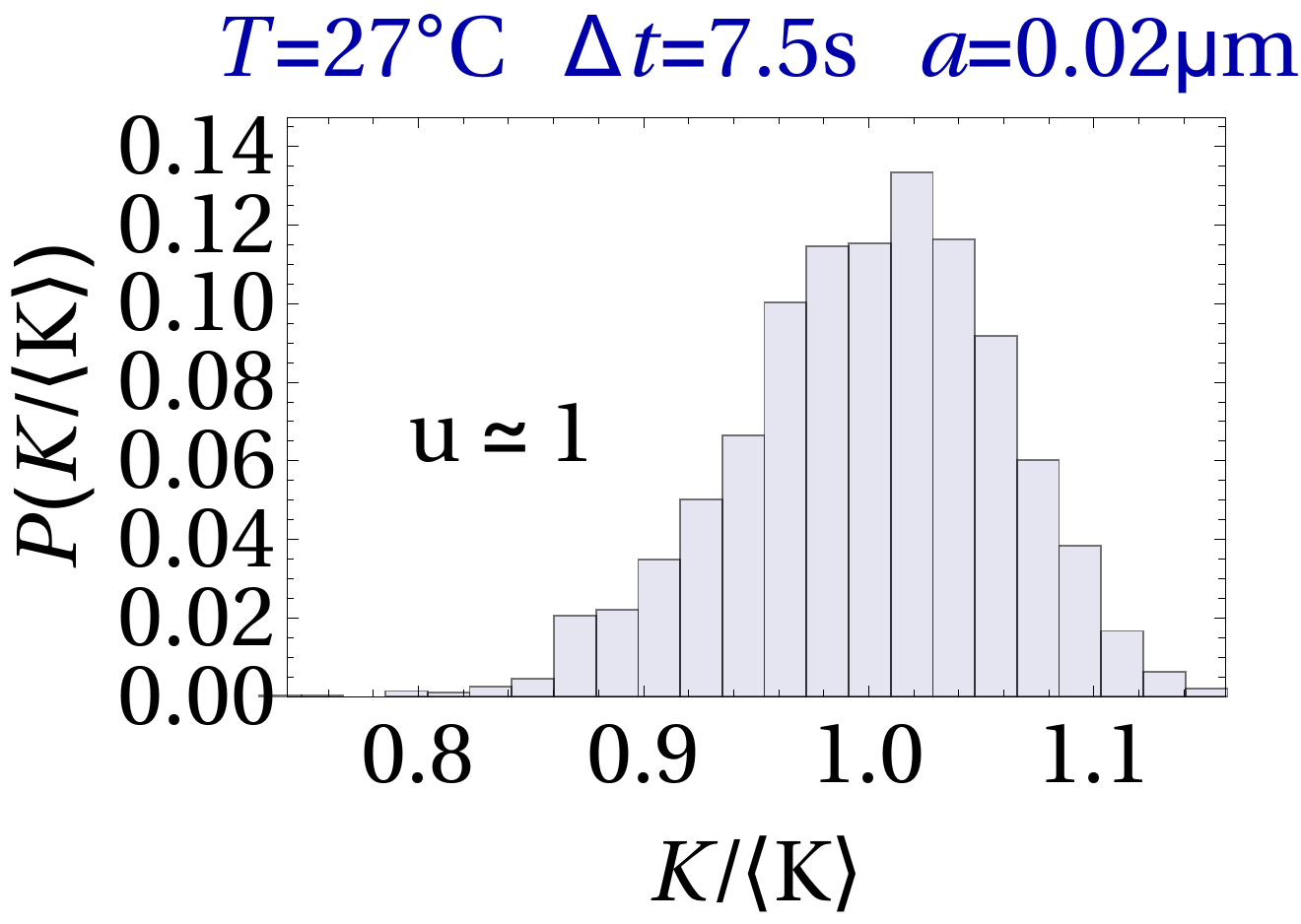}\hfill
\includegraphics[height=.23\columnwidth]{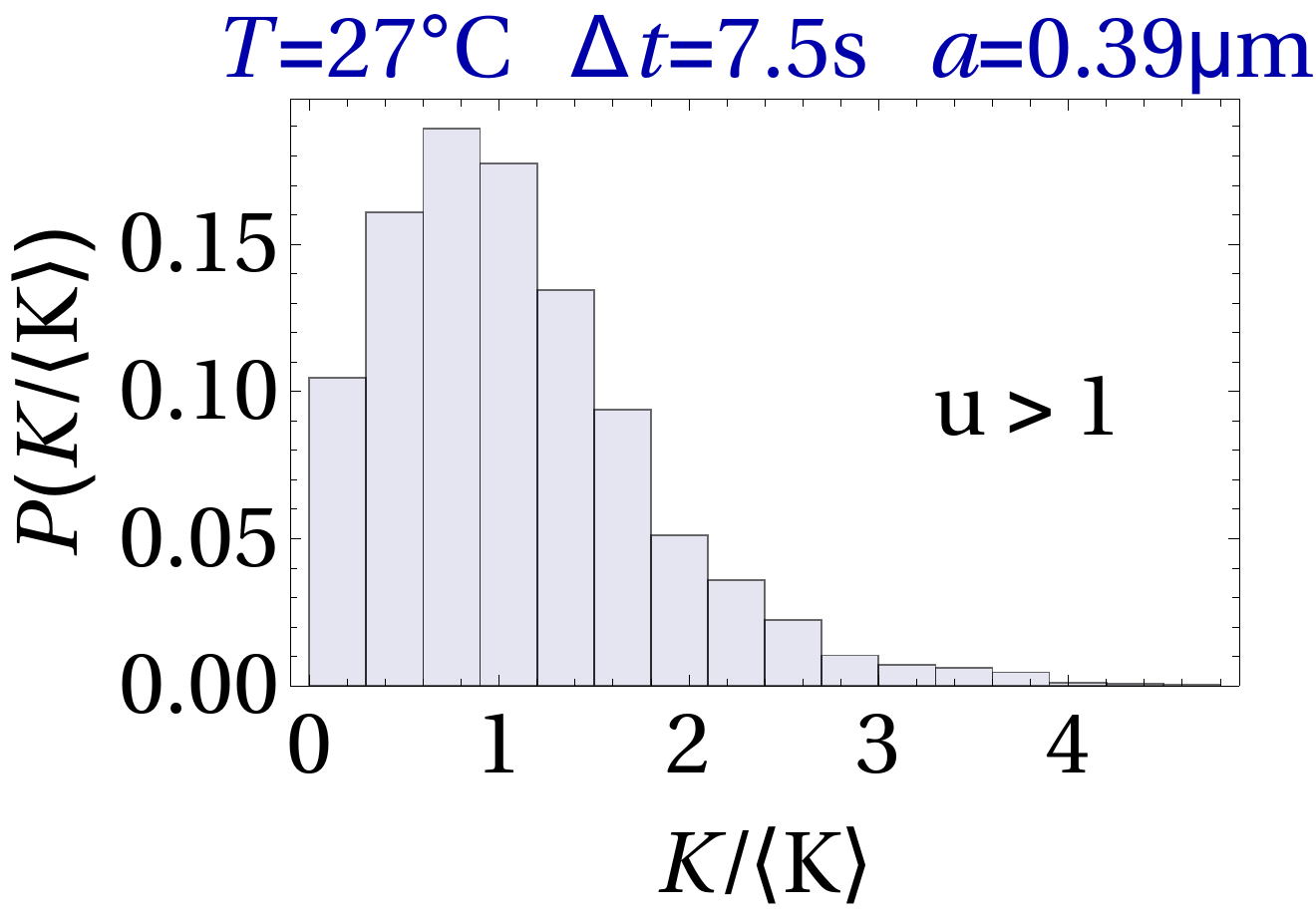}
\caption{From left to right, histogram of the activity at $T=27^\circ\text{C}$ normalized by its average for $a=0.021\sigma$, $a=0.085\sigma$ and $a=0.41\sigma$, respectively.
The corresponding values of $u$ are then respectively: 0.37, 1.50 and 7.35, and those of $\langle K \rangle$: 32.8, 24.1 and 4.71\,.
The time lapse $\Delta t$ is $7.5\,\text{s}$. The most symmetric histogram (center) is at the value of $a$ of the same order as $\Sigma(\Delta t)$.
   \label{hist27}}
\end{figure}

In Fig.~\ref{hist29} and~\ref{hist27}, we show the histograms of
$K/\langle K\rangle$ at both temperatures for varying values of $u$ --
smaller than, of the order of, and larger than $1$. For $u\sim1$
($a\sim\Sigma(\Delta t)$, \emph{i.e.}~for the most probable value of
the displacement at fixed $\Delta t$), the distribution of activity is
found to be symmetric, while it exhibits a pronounced
asymmetry when $a$ departs from $\Sigma(\Delta t)$ ($u<1$ and $u>1$). Besides, at fixed time
lapse $\Delta t$, the activity $K$ probes the distribution of jumps
larger than $a$. As a consequence, if $a\lesssim\Sigma(\Delta t)$, one
expects $K$ to be larger than if $a\gtrsim\Sigma(\Delta t)$: this is
indeed what we observe in Fig.~\ref{activityRemy}.

In a system with dynamical heterogeneities where slow and fast
trajectories coexist, $P(K)$ will exhibit two peaks, of small
and large activity, or at least an asymmetric shape biased towards
small $K$, especially if the peaks are not well separated. As the system
becomes more and more glassy, the width of $P(K)$, as well as the
asymmetry of $P(K)$, is expected to increase. In the two following
paragraphs, we will then focus on the variance and the skewness of the
activity with varying $u$ to investigate such a behavior, and detect the emergence of slow trajectories (low activity) for $u>1$.

% ranging from half a diameter up to three diameters.
%%
%% from /home/lecomtev/recherche/experiments_berangere-remy/donnees-remy_analysis-december-2014/analyse-resultats-remy_activite-statistique-T29_dec2014_final.nb
%%
%
%
%
%
%
%
%
%%
%% from /home/lecomtev/recherche/experiments_berangere-remy/donnees-remy_analysis-december-2014/analyse-resultats-remy_activite-statistique-T27_dec2014_final.nb
%%
%As is visible in Fig.~\ref{hist29}, for $a\sim\Sigma(\Delta t)$ (\emph{i.e.}~for the most probable value of the displacement at fixed $\Delta t$), the distribution of activity is symmetric. The histograms feature a pronounced asymmetry as $a$ departs from $\Sigma(\Delta t)$.
%we end up with a histogram that closely resembles a delta centered around 0. 
%This suggests that we should try and quantify the degree of asymmetry; this will be carried out in the next subsection by determining the skewness of the distribution.\\

%we end up with a histogram that closely resembles a delta centered around 0. 

%This suggests that we should try and quantify the degree of asymmetry; this will be carried out in the next subsection by determining the skewness of the distribution.\\
%%
%% from /home/lecomtev/recherche/experiments_berangere-remy/donnees-remy_analysis-december-2014/analyse-resultats-remy_activite-statistique-T27_dec2014.nb
%%
%ATTENTION! EST-CE K NORMALISE OU PAS???FAIRE UN CHOIX. IL EST PREFERABLE DE BIEN AVOIR ICI LA PDF DU K NORMALISE.
%ranging from a quarter of a diameter up to two diameters. 
%
Unlike what occurs at $T=29^\circ\text{C}$, asymmetry is slightly more marked at $a\neq\Sigma(\Delta t)$ and this is what we will quantify in the next subsection. This is visible with the naked eye only at $u<1$.

%the range of values $a\gtrsim\Sigma(\Delta t)$ is the most sensitive to dynamical heterogeneities, since those affect the regime of lower values of~$K$.
%

%Beside, as the system becomes more and more glassy (in our case, as temperature decreases), we expect the asymmetry of $P(K)$ to increase.

\subsection{Variance of the activity}

\begin{figure}[h]
\centering
\includegraphics[width=.75\columnwidth]{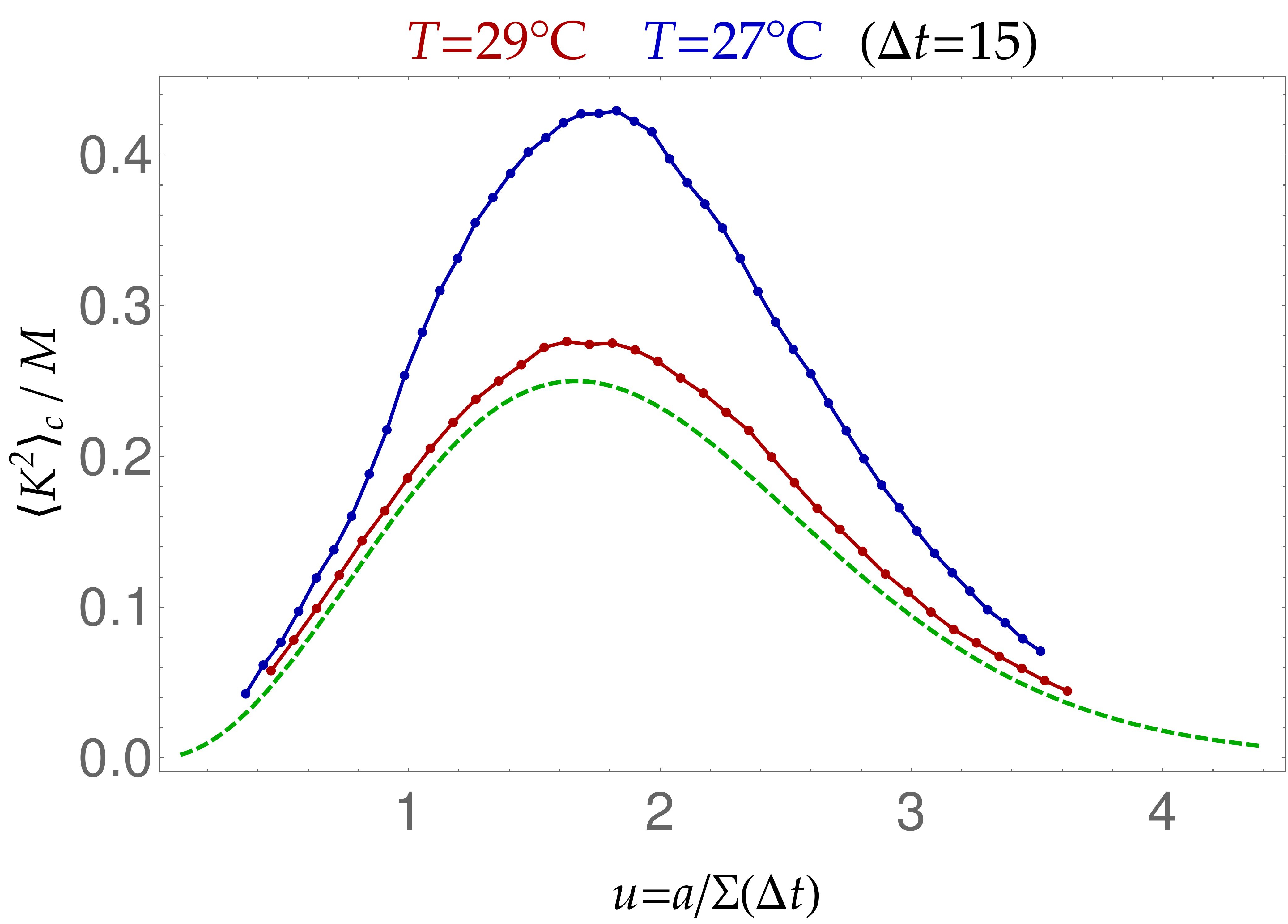}
\caption{Mean variance $\langle K^2\rangle_\cc/M$ as a function of
  $u={a}/\Sigma(\Delta t)$ at temperatures $T=29^\circ\text{C}$ (red)
  and $T=27^\circ\text{C}$ (blue) for the same time lapse $\Delta
  t=15\,\text{s}$. The dashed green curve stands for the Brownian
  motion counterpart Eq.~\eqref{eq:varianceBrownian} (in dimension
  $d=2$). 
   \label{varRemy1}
}
\end{figure}
%%
%% PLEASE DON'T REMOVE:
%%
%% from /home/lecomtev/recherche/analyse-resultats-remy_activite-statistique-T29_jan2015_TEST-math-4.nb
%%
%%
%
\begin{figure}[h]
\centering
\includegraphics[width=.49\columnwidth]{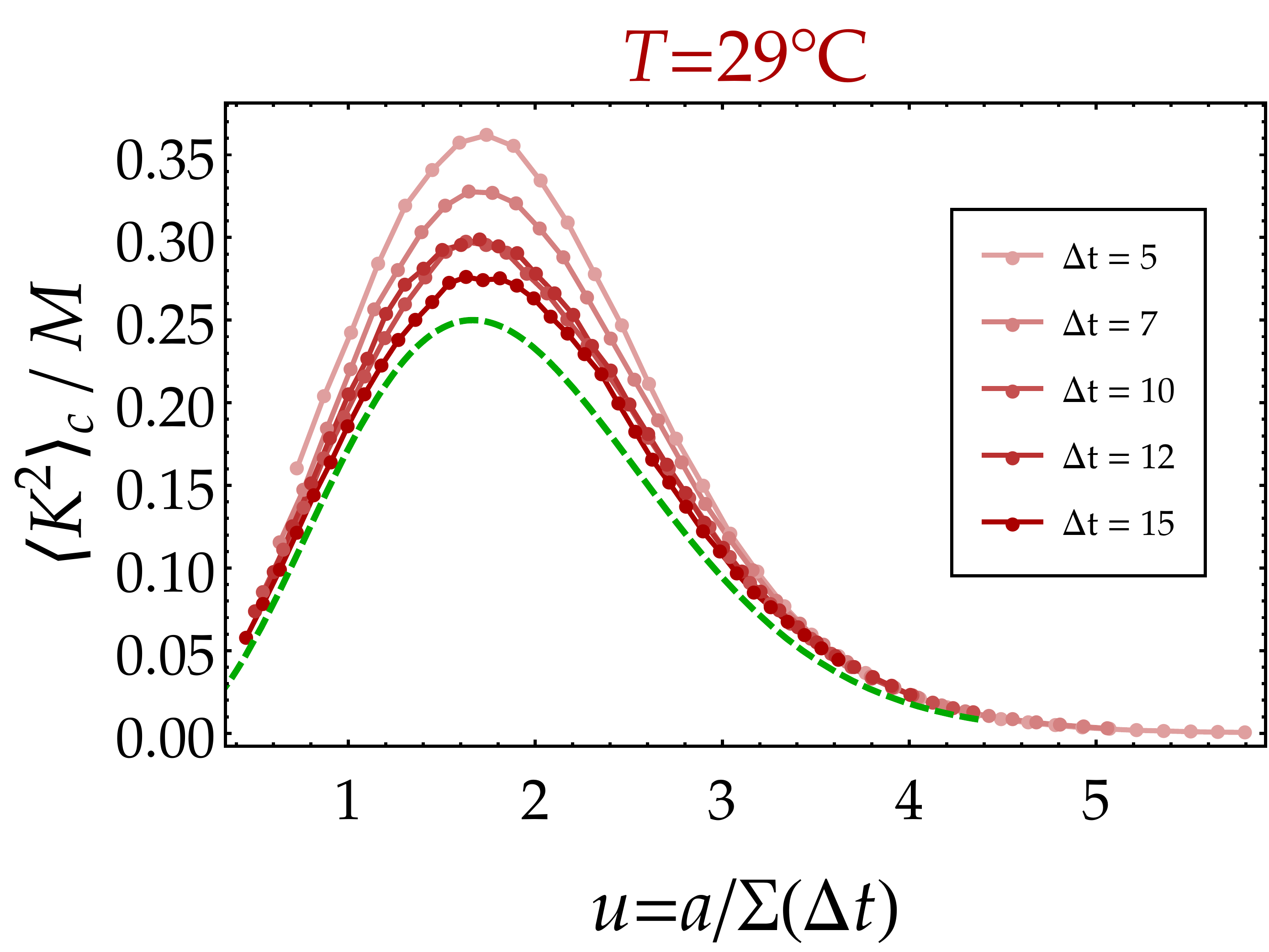}
\includegraphics[width=.49\columnwidth]{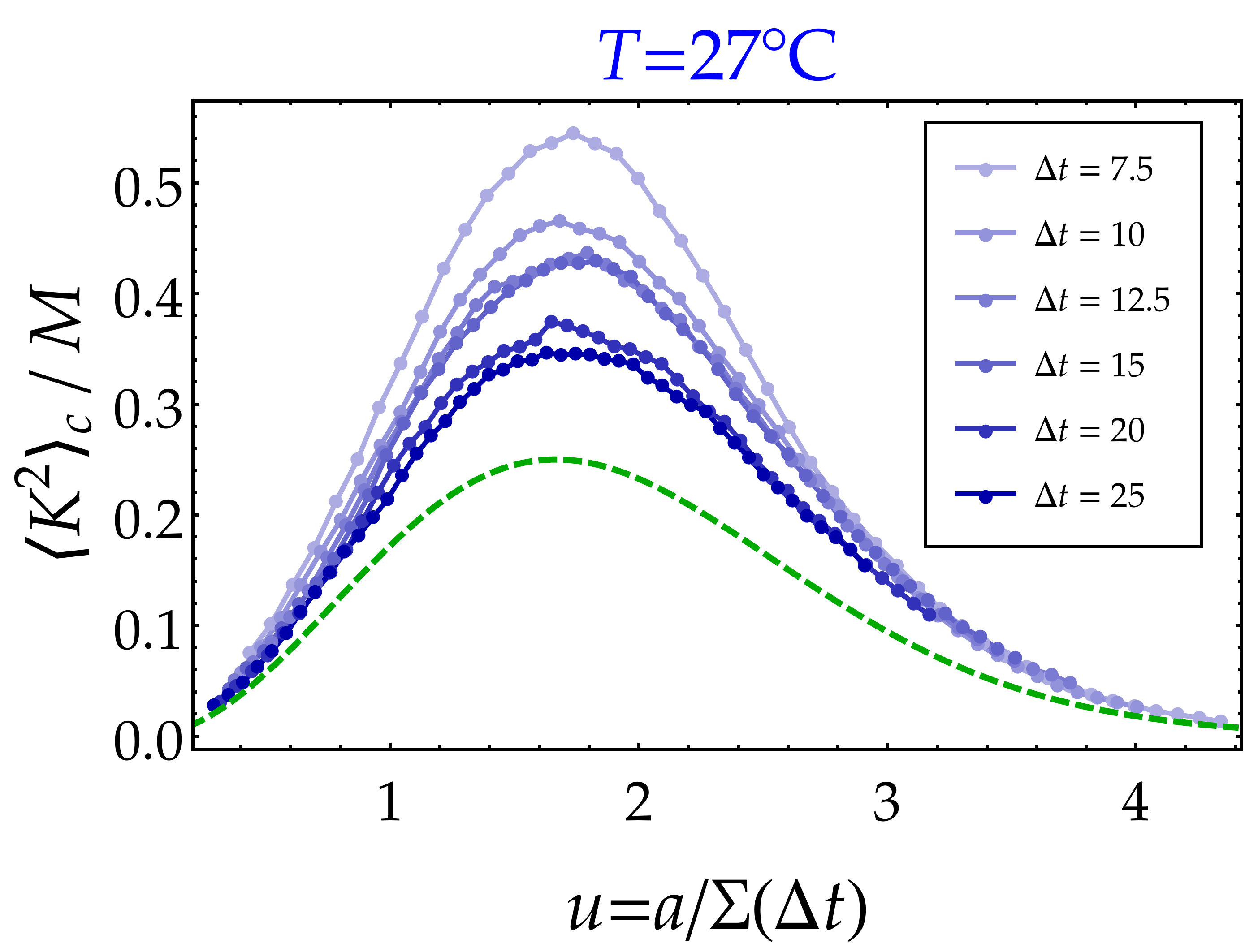}
\caption{Mean variance $\langle K^2\rangle_\cc/M$ as a function of $u={a}/\Sigma(\Delta t)$ at temperature $T=29^\circ\text{C}$ (\textbf{left}) and temperature $T=27^\circ\text{C}$ (\textbf{right}). The dashed green curves stand for the Brownian motion counterpart,
Eq.~\eqref{eq:varianceBrownian} (in dimension $d=2$). The various
curves correspond to different choices of $\Delta t$ as given in the
legend (values in seconds).
   \label{varRemy2}
}
\end{figure}%
%%
%% PLEASE DON'T REMOVE:
%%
%% from /home/lecomtev/recherche/analyse-resultats-remy_activite-statistique-T29_jan2015_TEST-math-4.nb
%% from /home/lecomtev/recherche/experiments_berangere-remy/donnees-remy_analysis-december-2014/analyse-resultats-remy_activite-statistique-T27_dec2014_final.nb
%%
%%

%Before turning to a quantitative analysis of the level of asymmetry of these distributions of activity $P(K)$, we investigate the behavior of their variance.
%
We compare in Fig.~\ref{varRemy1} the rescaled variances $\langle
K^2\rangle_\cc/M$ of $P(K)$ in the dense suspension at temperatures
$T=27^\circ\text{C}$ and $T=29^\circ\text{C}$, and the variance in the
Brownian case (equation~\eqref{eq:varianceBrownian}), for the same
value of the time lapse $\Delta t=15$s. In Fig.~\ref{varRemy2}, the
variances at each temperature are also plotted for different values of
the time lapse~$\Delta t$. Although the experimental data can not be
directly compared for the same $\Delta t$ (except for $\Delta t =
15\,\text{s}$), the variance in the suspension at low temperature
$T=27^\circ\text{C}$ is found to be larger than at high temperature
$T=29^\circ\text{C}$ (closer to the Brownian one). These results show
that the variance increases in a significant way when approaching the
glassy regime, indicating $P(K)$ displays a significant broadening,
consistent with the increase of dynamical heterogeneities.

We now want
to find out whether the corresponding broadening of $P(K)$ is due to a
symmetric enlargement of the central peak or whether it is due to the
emergence of rare events (on either side of the average).

%
%Since the scaling variable is $a/\Sigma(\Delta t)$, this corresponds to probing smaller spatial scales.
%
% \begin{figure}[h]
% \centering
% \includegraphics[width=.9\columnwidth]{varianceK_rescaled-by-RMSD_comparison-T-27-Brownian_Dtall.pdf}
% \caption{Variance as a function of ${a}/\Sigma(\Delta t)$ at temperatures $T=27^\circ\text{C}$ (blue). The continuous green curve stands for the Brownian motion counterpart. The various curves correspond to different choices of $\Delta t$.
%    \label{varRemy3}}
% \end{figure}
%, signals a broadening of the distribution of activity. 

\subsection{Skewness of the activity}

\begin{figure}[t]
\centering
\includegraphics[width=.75\columnwidth]{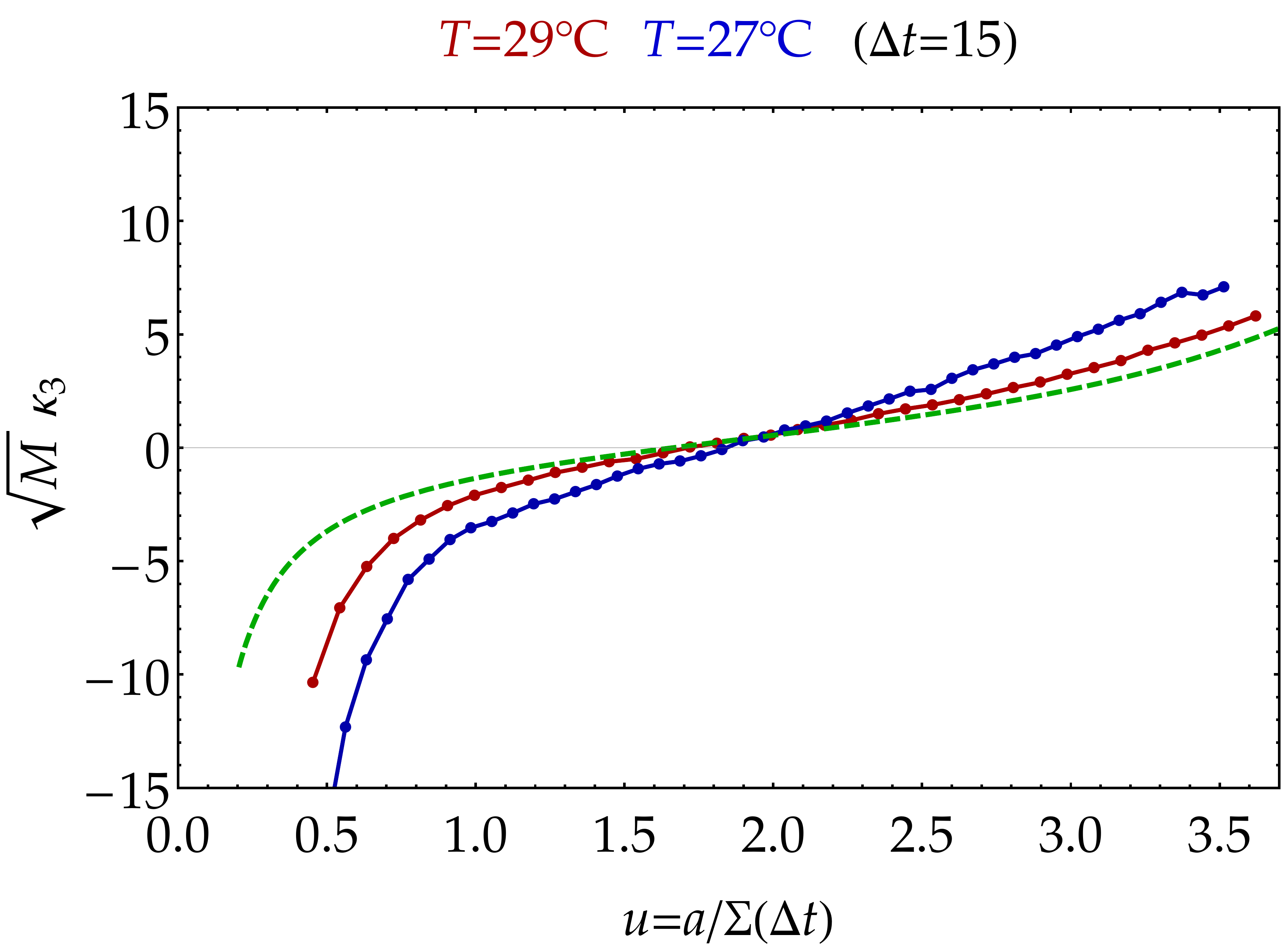}
\caption{Scaled skewness $\sqrt{M} \kappa_3$ as a function of $u={a}/\Sigma(\Delta t)$ at temperatures $T=29^\circ\text{C}$ (red) and $T=27^\circ\text{C}$ (purple) for the same time lapse $\Delta t=15\,\text{s}$. The dashed green curve stands for the Brownian motion counterpart Eq.~\eqref{eq:skewnessd2} in dimension $d=2$. 
   \label{skewRemy1}}
\end{figure}

\begin{figure}[t]
\centering
\includegraphics[width=.49\columnwidth]{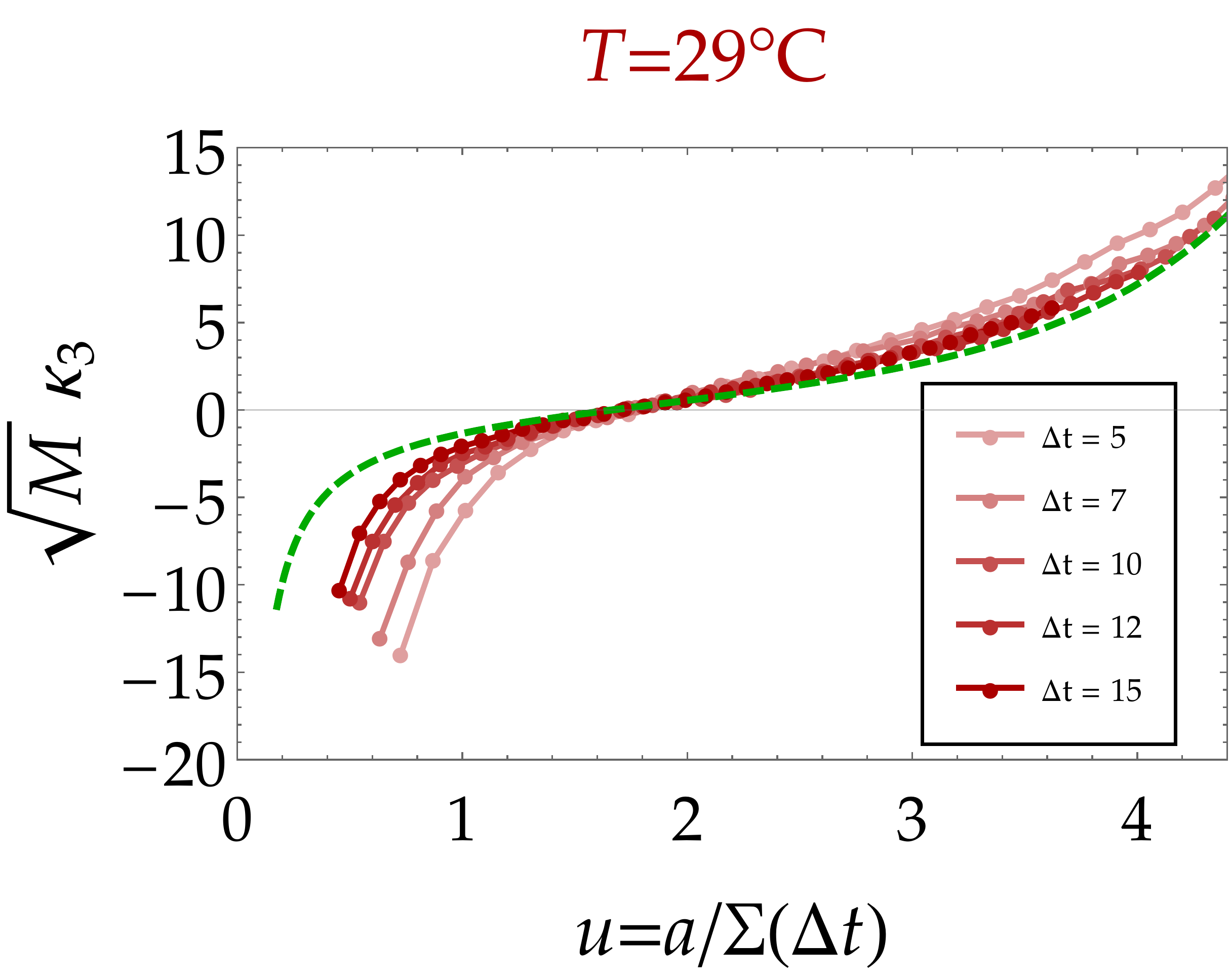}
\includegraphics[width=.49\columnwidth]{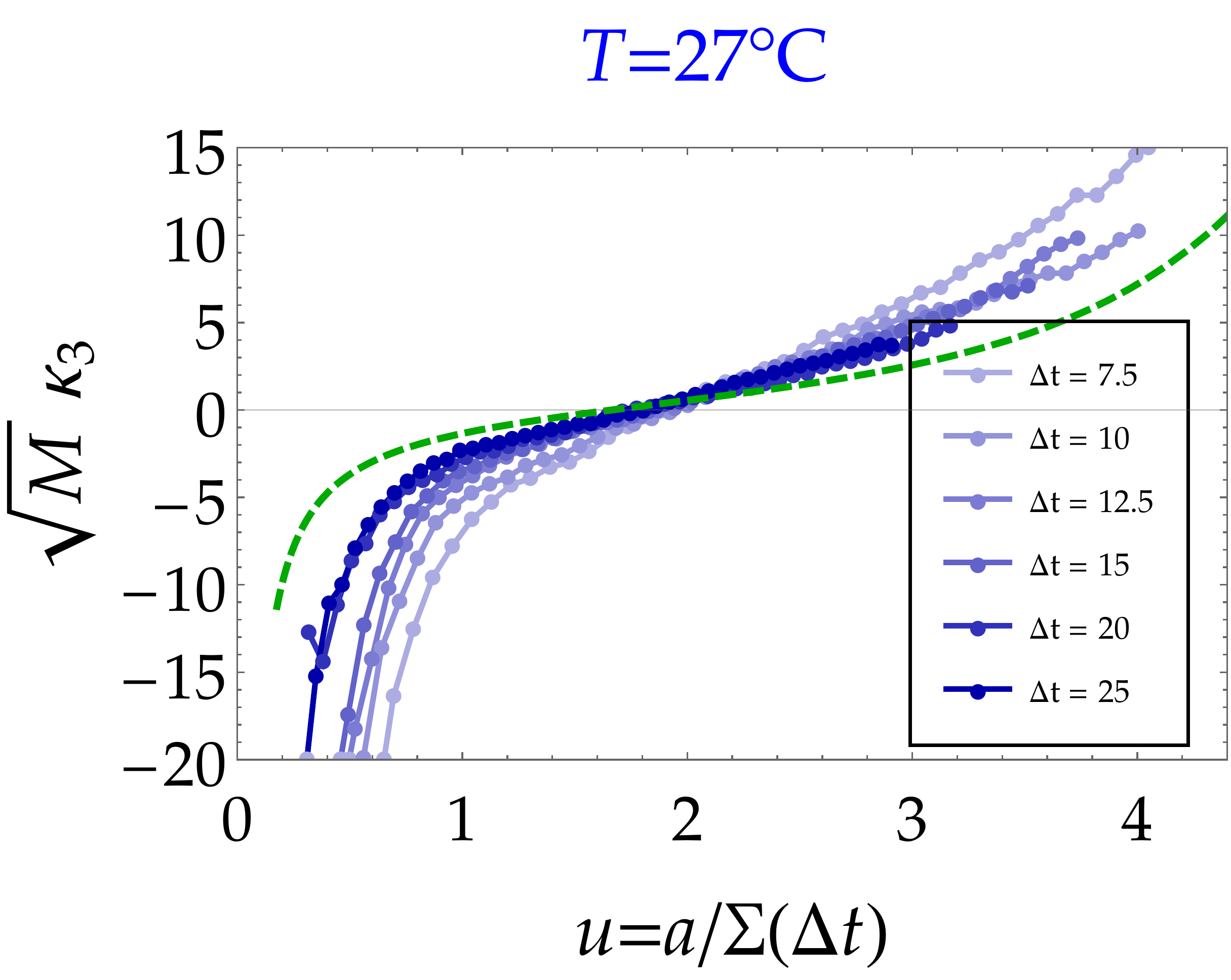}
\caption{Scaled skewness $\kappa_3\sqrt{M}$ as a function of $u={a}/\Sigma(\Delta t)$ at temperature $T=29^\circ\text{C}$ (\textbf{left}) and temperature $T=27^\circ\text{C}$ (\textbf{right}).
The dashed green curves stands for the Brownian motion counterpart Eq.~\eqref{eq:skewnessd2}. 
The various curves correspond to different choices of $\Delta t$ (in seconds).
   \label{skewRemy2}}
\end{figure}

We now investigate the asymmetry of the histograms $P(K)$ by focusing
on their skewness. Fig.~\ref{skewRemy1} compares the skewness for both
temperatures, and for the two-dimensional Brownian case (equation
\eqref{eq:skewnessd2}), for the same value $\Delta t=15$ s.
Around $u^* \simeq 2$, the skewness is zero for all the curves displayed, in agreement with the
symmetric distributions observed in Figs~\ref{hist29} and~\ref{hist27}
for the same value $u^*$.
For lower values $u<u^*$ and larger values $u>u^*$, the skewness departs from zero indicating the distributions -- including the Brownian case -- become asymmetric. In particular, in the large $u$ regime ($u>u^*$) where slow
trajectories are probed, the asymmetry is found to be significantly larger
with decreasing temperature. Since the domains number is quantified by the
skewness amplitude, our results provides a clear experimental evidence of the presence of a larger number of low activity
domains present in the suspension when approaching the glassy
regime. This effect can also been seen in Fig.~\ref{skewRemy2} where the scaled
skewness is plotted for various values of $\Delta t$, for both
temperatures.

%For a value of $a$ of the order of $\Sigma(\Delta t)$, the skewness is equal to zero, which corresponds to the symmetry of the distribution described in Sec.~\ref{ssec:histogramsK}. 
%
%The results displayed on Fig.~\ref{skewRemy1} now allow to characterize quantitatively the value of this symmetry point and one observes that it is the same in the experimental data and in the Brownian case.
%

% 
%One observes that the value of $a/\Sigma(\Delta t)$ is independent of $\Delta t$ (COMPREND PAS???), and that again the departure from the Brownian skewness is more asymmetric in the lower-temperature case, proving the robustness of the experimental signature of dynamical heterogeneities that we put forward.

% \textcolor{red}{A FAIRE : courbes de skewness : une pour T=27 et T=29 avec le même $\Delta t=15$, et avec le brownien ; une avec tous les $\Delta t$ (à voir) à $T=29^\circ\text{C}$, et pareil à $T=27^\circ\text{C}$.}
%
%
% \begin{figure}[h]
% \centering
% \includegraphics[width=.9\columnwidth]{skewnessK_rescaled-by-RMSD_comparison-T27-T29-Brownian.pdf}
% \caption{Skewness $\kappa_3\sqrt{M}$ as a function of $\frac{a}\Sigma(\Delta t)$ at temperatures $T=29$ (red) and $T=27$ (blue). The continuous green curve stands for the Brownian motion counterpart. For each temperature/color, the various curves corrrespond to different choices of $\Delta t$. At $T=27$ the $\Delta t$'s are such that $\sqrt{\langle (\br(t+\Delta t)-\br(t))^2\rangle}/\sigma$ lies between $0.XX$ and $0.XX$; and between $0.XX$ and $0.04$ at $T=29$.
%    \label{skewRemy}}
% \end{figure}

%%
%% PLEASE DON'T REMOVE:
%%
%% from /home/lecomtev/recherche/analyse-resultats-remy_activite-statistique-T29_jan2015_TEST-math-4.nb
%%
%%

\section{Discussion}

We have  put forward the experimental utility of a dynamical observable, the activity $K$, in order to quantify the approach to the glassy regime in dense microgel suspensions.
In the most  heterogeneous phase, the
distribution $P(K)$ of activity displays a secondary structure at
small values of $K$. This is a signature of long-lived slower than
average regions in the system.
Our main result is that, even far from the glass, in the supercooled regime, one observes  the existence of heterogeneities in the statistics of $K$, all the more so as one approaches the glass transition.
% in agreement with previous observation~\cite{colin_spatially_2011}

%}\DIFaddend .
%
We have compared the experimental results for $P(K)$ to the analytical results for a model of independent Brownian diffusers. 
The average of $K$ (which is fully determined by the static properties of the system), as expected, does not allow to distinguish the experimental tracers from the Brownian diffusers.
The mean variance of $K$ does display a difference: it increases as we go deeper in the glassy phase (by decreasing temperature), but it does not allow to distinguish between an effective broadening of the distribution of $K$ and the emergence of a dynamical phase with lower values of the activity.

The skewness of $K$, in contrast, proves to be the most relevant statistical observable, when measured as a function of the scaling variable $u=a/\Sigma(t)$.
%About the skewness we conclude:
%\begin{itemize}
%\item 
Using the diffusive ideal gas as a reference allows us to endow the deviations from the diffusive ideal gas with the meaning of effective number of {\it independent} degrees of freedom. At $T=29^\circ\text{C}$ (the liquid-like sample), we see that this number does not notably vary with the choice of $\Delta t$, for the whole $u\geq 1$ range. By contrast, at $T=27^\circ\text{C}$ (deeper in the supercooled liquid state hence more heterogeneous), we do see a greater sensitivity with respect to $\Delta t$ over the while $u$ range. 

% !!!!!ATTENTION A LA DISCUSSION CRR!!!THIS IS MISLEADING!!!

%This qualitatively echoes the findings of Crauste-Thibierge {\it et al.}~\cite{PhysRevLett.104.165703}. The tracers in the least glassy sample behave almost like a diffusive ideal gas in terms of the activity, independently of the window scale: there is dynamical homogeneity. Dynamical heterogeneity appears to be stronger at $T=27^\circ\text{C}$ by a decrease in the effective number of dynamical degrees of freedom.
%
%\item 

The distribution of activity for Brownian particles is obviously asymmetric (because it probes atypical events which have no reason to be Gaussian). A positive skewness indicates an excess of larger-than-average events. For $u\geq 1$, not only do we have a positive skewness, but above all the latter is in large excess over the Brownian curve. In the glassy state, there is an excess of longer range directed events.
% which we view as being consistent with the increase of the size of the Cooperatively Rearranging Regions~\cite{AdamGibbs-1st}.

%\item 
Our interest goes now to the skewness falling below the Brownian level at values of $u\leq 1$. In this $u$ regime ($u$ a fraction of unity) we know that $a$ really has the meaning of a cage size. And we see that there is a sharp increase in less-than-average active events. This observation is consistent with the emergence of a secondary low activity peak in the activity distribution, without having to characterize the large deviations of $P(K)$ (which are difficult to measure in experiments). What we witness here is the build-up of inactive events that leads to a fatter-than-Brownian tails in inactive events.

%\end{itemize}
In Appendix~\ref{app:weeks}, we show that our results are robust and fully consistent with what can be inferred from experimental data by Weeks {\it et al.}~\cite{WeeksScience}. This comparison
illustrates the robustness of our proposed analysis, which still holds although the data of Ref.~[\onlinecite{WeeksScience}] present the following differences: (\emph{i}) the tracking is performed in dimension $d=3$ instead of our effective $d=2$, (\emph{ii}) instead of specific tracers, all particles of the system are tracked, and (\emph{iii}) the acquisition is made on shorter trajectories in time, but with larger statistics. 

\noindent {\bf Acknowledgements:} We warmly thank Eric Weeks for allowing us to make use of his data and for his comments on an earlier version of this manuscript.
VL acknowledges support by the the ERC Starting Grant 680275 MALIG and by the ANR-15-CE40-0020-03 Grant LSD.

\appendix
\section{Appendix: Microgel suspension behavior with volume fraction}
\label{app:pNipam}
\begin{figure}[tbp]
\begin{center}
\includegraphics[width=\columnwidth]{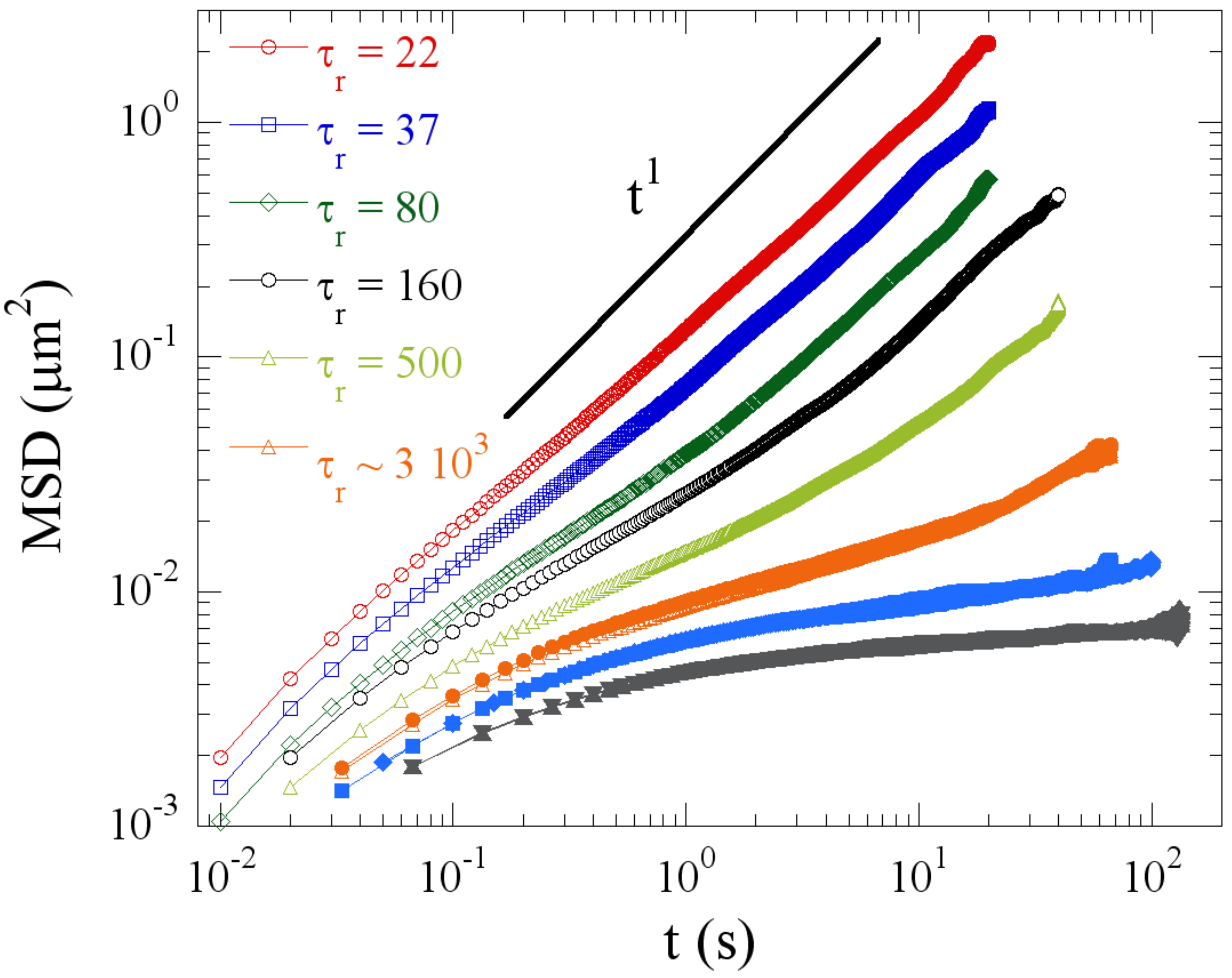}
\end{center}
\caption{Mean-squared displacement of Latex tracers ($0.994 \,\mu$m in diameter) as
 a function of the lag time measured in the 
microgel suspensions. With increasing volume fraction, 
 we observe a short-time diffusive behavior, then a sub-diffusive behavior 
 at intermediate time scales. These two regimes are followed by a long-time 
 diffusive behavior. At the highest volume fraction, a plateau  
 develops and the crossover to the long-time 
 diffusive regime could definitely not be reached within reasonable 
 experimental timescales. A relative relaxation time $\tau_r$ is deduced
 from the long-time diffusion coefficient when measured. }
\label{fig:Remy6_1}
\end{figure}

%
%
%First figure (figure 6.1 these Remy)
%protocol and MSD description
Our purpose here is to describe the phase behavior of our microgel
suspensions, based on parameters such as a relative relaxation time or
an effective volume fraction. Figure~\ref{fig:Remy6_1} shows the MSD
of Latex tracers ($0.994 \,\mu$m in diameter) in the microgel suspension
at various volume fractions. This latter parameter was increased in a
quasistatic way, by performing temperature incremental step
increases. The suspension was allowed to relax between each step to
reach an equilibrium state. At low volume fraction, the suspension is
in a liquid state as characterized by the linear dependency of the MSD
with the lag time. Upon increasing volume fraction, the MSD typically
exhibits a short-time diffusive regime, followed by a sub-diffusive
regime at intermediate timescales, and again a long-time diffusive
regime, which can only be measured when the suspension is not too deep
in the supercooled states. At the highest volume fractions, a plateau
develops and the crossover to the long-time diffusive regime could
definitely not be reached within reasonable experimental timescales.
 
 %determination of effective volume fraction and relative relaxation time
 %
 From the mean-squared displacement data, one can infer a relative relaxation 
 time $\tau_r$ and an effective volume fraction $\Phi_{\rm eff}$ which calculation is already 
 described in Ref.~[\onlinecite{Questioning2015}]. The relative relaxation 
 time $\tau_r$ was deduced from the long-time diffusion coefficient $D_\infty$, 
 with $\tau_r=\tau(T)/\tau_0(T)=\eta(T)/ \eta_0 (T)= D_0/D_\infty $, where $\tau, \eta, D_\infty$ and 
 $\tau_0, \eta_0, D_0$ are respectively the 
 probe diffusion time, viscosity, and long-time diffusion coefficient in the 
  microgel suspension and in water. 
%Both relative relaxation times and effective 
%  volume fractions follow a master curve shown in Fig.~\ref{fig:Remy6_2}, giving evidence 
%  of the equivalence 
%  between both parameters. The relative relaxation time, or 
%  relative viscosity $\tau_r=\eta/ \eta_0$, increases with volume fraction. For low volume 
%  fractions between $0.4 < \Phi_{\rm eff}<0.5$, the data  
% obtained in 
% hard sphere suspensions by Meeker et al.~\cite{Meeker1997} 
% provides a good description of our data. Around 
%  effective volume fraction $\Phi_{\rm eff}=0.66-0.70$, the increase is sharp, giving an estimate 
%  of the glass transition 
% volume fraction.  

We have previously shown that the signature of dynamical heterogeneities 
that characterize the supercooled regime were 
encapsulated in how the PDFs deviate from a Gaussian~\cite{colin_suspensions_2012, 
 colin_spatially_2011}. Based on their dynamical properties, our suspensions could be classified, 
from liquids, to supercooled liquids and finally to glasses, when aging occurs on experimental timescales. The curves presented in this study, 
with large relaxation times ${\tau_r}_{29}=128\pm 7$ and ${\tau_r}_{27}=649\pm 32$, and non Gaussian PDFs, are found to be 
in the supercooled liquid state. 
%(linear MSD, Gaussian PDF, fast relaxation time,), 
%(linear or non linear MSD, non Gaussian PDF, slower relaxation time), to glasses 
%(where aging occurs  ).Estelle j ai un souci avec la classification on peut en discute par tel
 
%
%Second figure (figure 6.2 these Remy)
%When the onset of diffusive regime is at larger time than the observation time, 
%no relaxation time can be inferred. However, one can obtain an estimation of the 
%approach of the glass transition: how $phi_g$ is induced?

% \begin{figure}[tbp]
% \begin{center}
% \includegraphics[width=\columnwidth]{Figure6_2_these_Remy.pdf}
% \end{center}
% \caption{Relative relaxation time as a function of an effective 
% volume fraction in our microgel suspensions ($\circ$,$\blacksquare$). 
% The straight line is obtained 
% in hard spheres suspensions, 
% by Meeker {\it et al.}~\cite{Meeker1997} and provides a good description of our soft sphere
% suspensions between $0.4 < \Phi_{\rm eff}<0.5$. Around 
%  an effective volume fraction $\Phi_{\rm eff}=0.66-0.70$, the increase is sharp, giving an estimate 
%  of the glass transition 
% volume fraction. }
% \label{fig:Remy6_2}
% \end{figure}

\section{Appendix: Activity for a Brownian motion}
\label{app:A_BM}
Our Brownian particle has diffusion constant $D$. We ask how the
activity of a given particle with trajectory $\br(t)$ is
distributed. The generating function $\hat{P}(s,t)=\langle\ee^{-s
  K(t)}\rangle$ of the activity is written as
$\hat{P}(s,t)=\langle\ee^{-s\sum_{m=0}^{M-1}\Theta(||\Delta\br_m-a)}\rangle$,
where $\Delta\br_m=\br((m+1)\Delta t)-\br(m\Delta t)$. Using that all
segments of the trajectory are independent, we end up with
$\hat{P}(s,t)=\langle\ee^{-s \Theta(||\Delta\br||-a)}\rangle^{M}$. The
argument in between the average brackets is unity if the excursion
$||\Delta\br||$ remains smaller than $a$ and $\ee^{-s}$ otherwise. As
defined, our activity is thus a positive number varying between $0$
and $M$. Given that the probability of a displacement
$\ell=||\Delta\br||$ is, in 3d, $p(\ell,\Delta
t)=\frac{4\pi\ell^2}{\sqrt{4\pi D \Delta
    t}^3}\ee^{-\frac{\ell^2}{4D\Delta t}}$ we arrive at
\begin{equation}
\hat{P}(s,t)=\left[\int_0^a\dd\ell p(\ell,\Delta t)+\ee^{-s}\int_a^\infty\dd\ell p(\ell,\Delta t)\right]^{M}
\end{equation}
Once the generating function $\hat{P}(s,t)$ is known one can reconstruct the full distribution by inverting the Laplace transform according to $P(K,t)=\int\frac{\dd s}{2\pi i}\ee^{s K}\hat{P}(s,t)$. At asymptotically large times, one can find the behavior of $P(K,t)$ to be given, in terms of $k=\frac{K}{M}$ and $u=\frac{a}{\sqrt{D\Delta t}}$, by
\begin{equation}\begin{split}
\frac{\ln P(K,t)}{ M}
\simeq &k\ln\left[\frac{\text{erfc}\left(\frac{u}{2}\right)-1+\frac{u\ee^{-\frac{u^2}{4}}}{\sqrt{\pi}}}{k-1}\right]\\
&+k\ln\left[\frac{k-1}{k}\frac{\frac{u\ee^{-\frac{u^2}{4}}}{\sqrt{\pi}} +\text{erfc}\left(\frac{u}{2}\right)}{\frac{u\ee^{-\frac{u^2}{4}}}{\sqrt{\pi}} -\text{erf}\left(\frac{u}{2}\right)}\right]
\end{split}
\end{equation}
The average activity maximizes $P(K,t)$ and it reads
\begin{equation}\label{meanact_app}
\frac{\langle K\rangle}{M}=\left\{\begin{array}{ll}
\text{erfc}\left(\frac{u}{2}\right)+\frac{\ee^{-\frac{u^2}{4}} u}{\sqrt{\pi }}&{(d=3)}\\
\ee^{-\frac{u^2}{4}}&{(d=2)}
\end{array}\right.
\end{equation}
and the right hand side in~\eqref{meanact_app} is bounded by 0 and 1. Asymptotic expressions for the higher moments or the cumulants of $K$ can be found by similar means. The variance is given by
\begin{equation}
\hspace*{-10mm}
\langle K^2\rangle_\cc=\left\{
\begin{array}{ll}
 M (\ee^{-\frac{u^2}{4}}-\ee^{-\frac{u^2}{2}})&\text{ if }d=2\\
\frac{M e^{-\frac{u^2}{2}} \left(\sqrt{\pi } \ee^{\frac{u^2}{4}}
   \text{erf}\left(\frac{u}{2}\right)-u\right) \left(\sqrt{\pi } \ee^{\frac{u^2}{4}}
   \text{erfc}\left(\frac{u}{2}\right)+u\right)}{\pi }&\text{ if }d=3
\end{array}
\right.
\label{eq:varianceBrownian}
\end{equation}
and  the first nontrivial signature of a deviation with respect to the Gaussian distribution, namely the normalized third cumulant, otherwise known as the skewness $\kappa_3$ of the distribution reads
\begin{small}%
\begin{equation}%
\hspace*{-12mm}
\kappa_3=\frac{1}{\sqrt{M}}\frac{\ee^{-\frac{u^2}{4}} \left(\sqrt{\pi } \ee^{\frac{u^2}{4}} \left(2
   \text{erf}\left(\frac{u}{2}\right)-1\right)-2
   u\right)}{\sqrt{\sqrt{\pi } \ee^{-\frac{u^2}{4}} u \left(2
   \text{erf}\left(\frac{u}{2}\right)-1\right)-\pi 
   \left(\text{erf}\left(\frac{u}{2}\right)-1\right)
   \text{erf}\left(\frac{u}{2}\right)-\ee^{-\frac{u^2}{2}} u^2}}
\end{equation}%
\end{small}%
A similar calculation carried out in space dimension 2 leads to 
\begin{equation}\begin{split}
\frac{\ln P(K,t)}{M}
\simeq& -\frac{k u^2}{4}+\ln \left(\frac{\ee^{-\frac{u^2}{4}}-1}{k-1}\right)\\&
+k \ln  \left(-\frac{(k-1) \left(\coth \left(\frac{u^2}{8}\right)+1\right)}{2 k}\right)
\end{split}
\end{equation}
along with 
\begin{equation}
\kappa_3=\frac{1}{\sqrt{M}}\frac{-2+\ee^{\frac{u^2}{4}}}{\sqrt{-1+\ee^{\frac{u^2}{4}}}}
\end{equation}

\section{Appendix: Results from Weeks data analysis}\label{meatWeeks}
\label{app:weeks}

\subsection{Description of the experimental system }

A number of measurements have been performed to quantify the motion of colloidal particles near the glass transition. We focus here on the simplest system, i.e. monodisperse colloids (poly-methylmethacrylate particles stabilized by a thin layer of poly-12-hydroxystearic acid), both in equilibrated supercooled colloids fluids and non-equilibrated glasses, as used in the pioneering work on dynamical heterogeneities by Weeks {\it et al.}~\cite{WeeksScience}. The particles have a radius $a_0=1.18 \mu m$.
We chose to analyze the 3d motion of particles recorded by confocal microscopy in two distinct systems. The first one is a supercooled fluid, with density $\phi=0.46$, the time step between consecutive images is $\delta t _{0.46}=10s$, the duration of the movie is $271 \delta t_{0.46}$, and the number of tracked particles is 4232. The second one is a glass studied after a long period of ageing, with density $\phi=0.6$, the time step between consecutive images is $\delta t _{0.6}=120s$, the duration of the movie is $333\delta t_{0.6}$, and the number of tracked particles is 5922. 
We have also analyzed the data concerning an intermediate system with density $\phi=0.52$, which is a denser but still supercooled fluid, with time step between consecutive images being  $\delta t _{0.52}=18s$, duration of the movie $431 \delta t_{0.52}$   and the number of tracked particles is 4679.  We will also present some of the results from this data set. 
We note that although the most glassy samples are ageing, this can be neglected over the duration of the measurements.
The tracking procedure is hence different from ours: here the coordinates of all particles were tracked, whereas in our experiments, the motion of a few tracers only
was recorded. Moreover, the time duration of the movies recorded in Weeks {\it et al.}~\cite{WeeksScience} is relatively short considering the displacements distribution function (the displacement do not exceed the particle radius $a_0$); whereas in our experiments, the movies are longer in the sense that the PDF of the displacements
samples distances far as $\sim 6$ times the particle radii $\sigma/2$ (this is compensated by higher tracer statistics). 
We computed the activity for different values of $a$ and $\Delta t=t_{obs}/M$. In the following $\Delta t$ is expressed in units of $\delta t$. Systems and tracking approaches differ, they will nevertheless be shown to point in the same direction.

\subsection{Histogram of activity}
\begin{figure}[tbp]
\begin{center}
\includegraphics[width=.75\columnwidth]{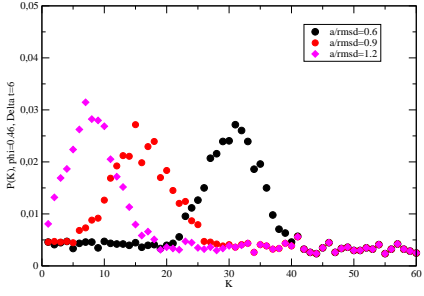}
\end{center}
\caption{Histogram of the activity $K$ at $\phi=0.46$ normalized by its average, $\Delta t=6 \delta t_{0.46}$, for three values of $u$ equal to 0.6, 0.9, 1.2 respectively. The histograms are symmetric.}
\label{fig:WeeksHisto0.46}
\end{figure}

\begin{figure}[tbp]
\begin{center}
\includegraphics[width=.75\columnwidth]{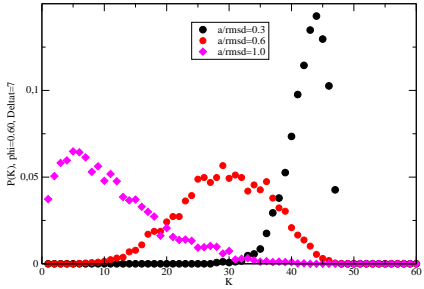}
\end{center}
\caption{Histogram of $K$ at $\phi=0.6$
normalized by its average, $\Delta t=7\delta t_{0.6}$ for three values of $u$ equal to 0.3, 0.6, 1 respectively. 
The scaled skewness represented on~\ref{fig:WeeksSkew0.46_0.6} provides one a quantitative indicator that the sample at $\phi=0.6$ is clearly heterogeneous, compared to the sample at $\phi=0.46$, because of its clear departure from the purely Brownian behavior.
%
% The histograms are symmetric for the two smaller values of $a$. By contrast with the previous case, visual inspection shows clearly that the histogram for the larger value of $a$ is asymmetric and displays an excess tail in larger than average activity events.
}
   \label{fig:WeeksHisto0.6}
\end{figure}

\begin{figure}[tbp]
\begin{center}
\includegraphics[width=.75\columnwidth]{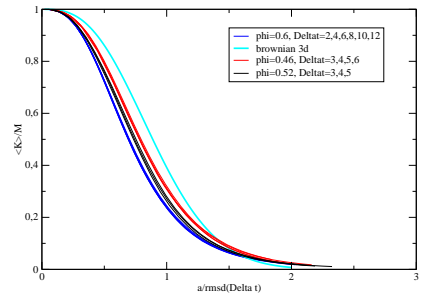}
\end{center}
\caption{Mean activity $\langle K\rangle/M$ at $\phi=0.46$, $0.52$ and $0.6$ versus the Brownian result~\eqref{meanact} (for $d=3$ dimensions). For each density $\phi$, $\Delta t$s are expressed in units of $\delta t_{\phi}$. The mean activity decreases only slightly as a function of density.}
   \label{fig:WeeksMeanKvsBrownian}
\end{figure}

In Fig.~\ref{fig:WeeksHisto0.46} and Fig.~\ref{fig:WeeksHisto0.6}, we represent the distribution of activity for the two particle densities.
The denser (and more heterogeneous) system displays the strongest asymmetry when varying $a/\Sigma(\Delta t)$.
In Fig.~\ref{fig:WeeksMeanKvsBrownian}, we illustrate that the average of $K$ does not allow to distinguish dynamically and non-dynamically heterogeneous situations, in the same way as for our experimental data (see Fig.~\ref{activityRemy}).
The observed value for the average activity  is very close to the analytical result~\eqref{meanact} for Brownian particles in three dimensions.

\subsection{Average variance of the activity}

\begin{figure}[htbp]
\begin{center}
\includegraphics[width=.75\columnwidth]{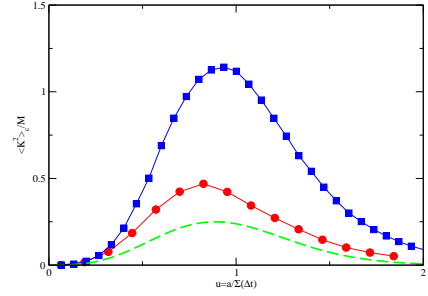}
\end{center}
\caption{Mean variance $\langle K^2\rangle_\cc/M$ as a function of $u$ at $\phi=0.46$ and $0.52$ for the same $\Delta t=9 \delta t_{0.46}=5\delta t_{0.52}= 90s$.}
   \label{fig:WeeksVar0.46-0.52}
\end{figure}

\begin{figure}[htbp]
\begin{center}
\includegraphics[width=.49\columnwidth]{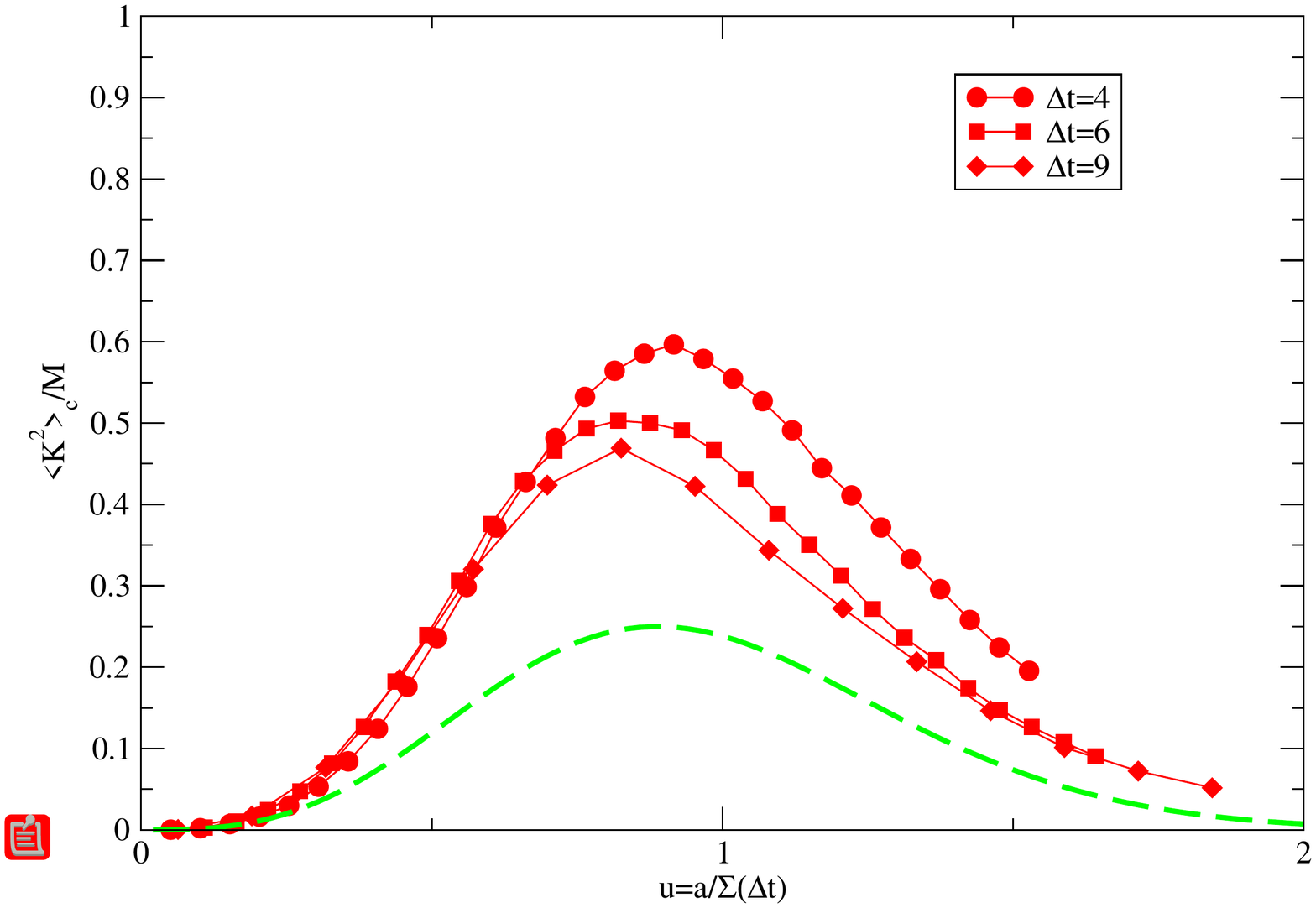}
\includegraphics[width=.49\columnwidth]{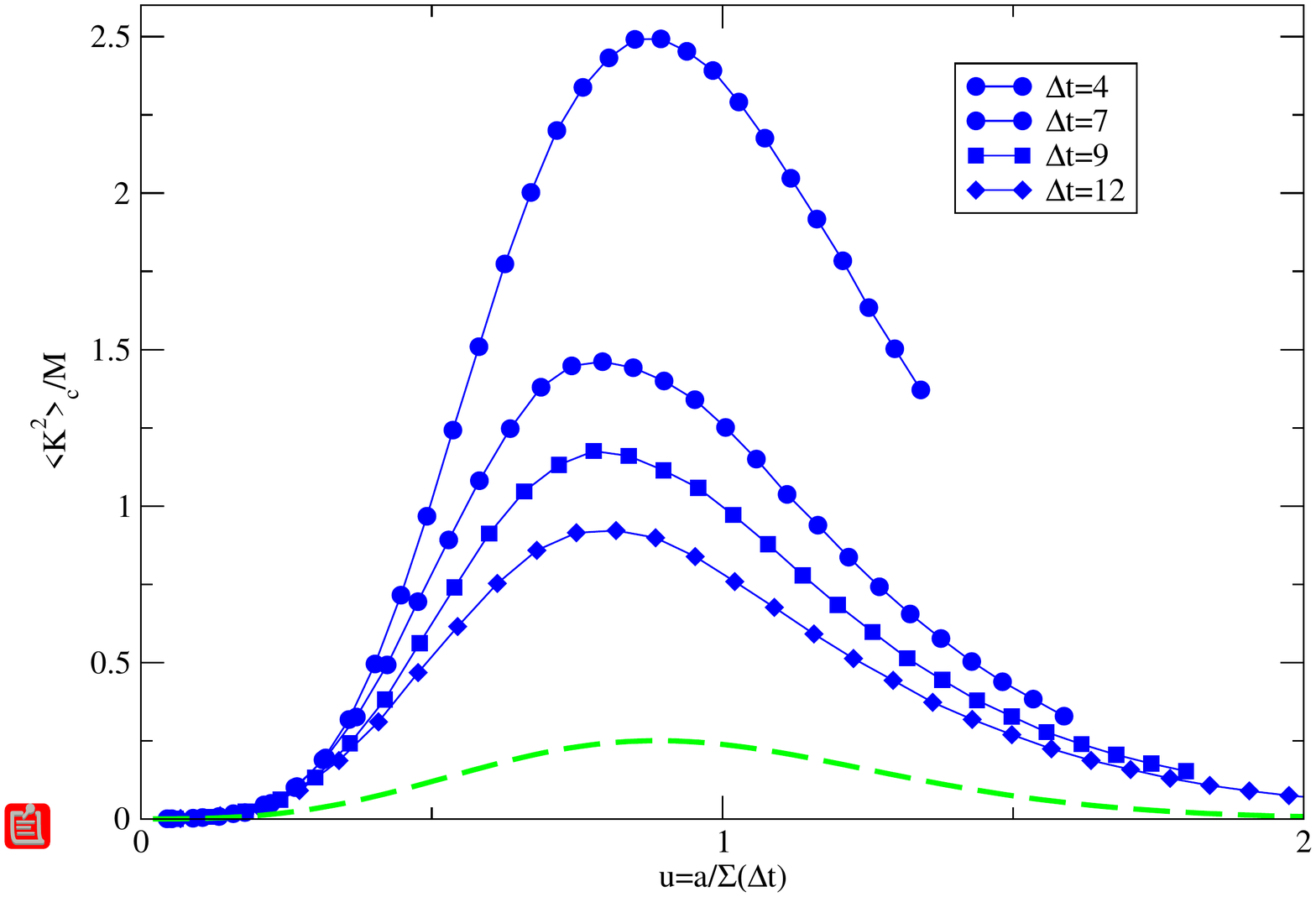}
\end{center}
\caption{Mean variance $\langle K^2\rangle_\cc/M$ as a function of $u$ for the 2 densities $\phi=0.46$ and $\phi=0.6$ and different values of $\Delta t$ expressed as units of $\delta t_{\phi}$. The Brownian curve is plotted as a reference.}
   \label{fig:WeeksVar0.6_0.46}
\end{figure}

In Fig.~\ref{fig:WeeksVar0.46-0.52} and Fig.~\ref{fig:WeeksVar0.6_0.46}, 
we illustrate (as for our experimental data in~Fig.~\ref{varRemy1} and~Fig.~\ref{varRemy2}) that more glassy (denser) system display an increase in the variance.
The Brownian result~\eqref{eq:varianceBrownian} (in dimension $d=3$) serves as a comparison, and is much smaller in all cases.

% \begin{figure}[htbp]
% \begin{center}
% \includegraphics[width=.75\columnwidth]{variance-VarK_Mscaled_a_over_rmsd_phi060.pdf}
% \end{center}
% \caption{Mean variance $\langle K^2\rangle_\cc/M$ at $\phi=0.6$.}
%    \label{fig:WeeksVar0.6}
% \end{figure}

% \begin{figure}[htbp]
% \begin{center}
% \includegraphics[width=.75\columnwidth]{variance-VarK_Mscaled_a_over_rmsd_phi046.pdf}
% \end{center}
% \caption{Mean variance $\langle K^2\rangle_\cc/M$ at $\phi=0.46$.}
%    \label{fig:WeeksVar0.46}
% \end{figure}

\subsection{Scaled skewness of the activity}

\begin{figure}[htbp]
\begin{center}
\includegraphics[width=.49\columnwidth]{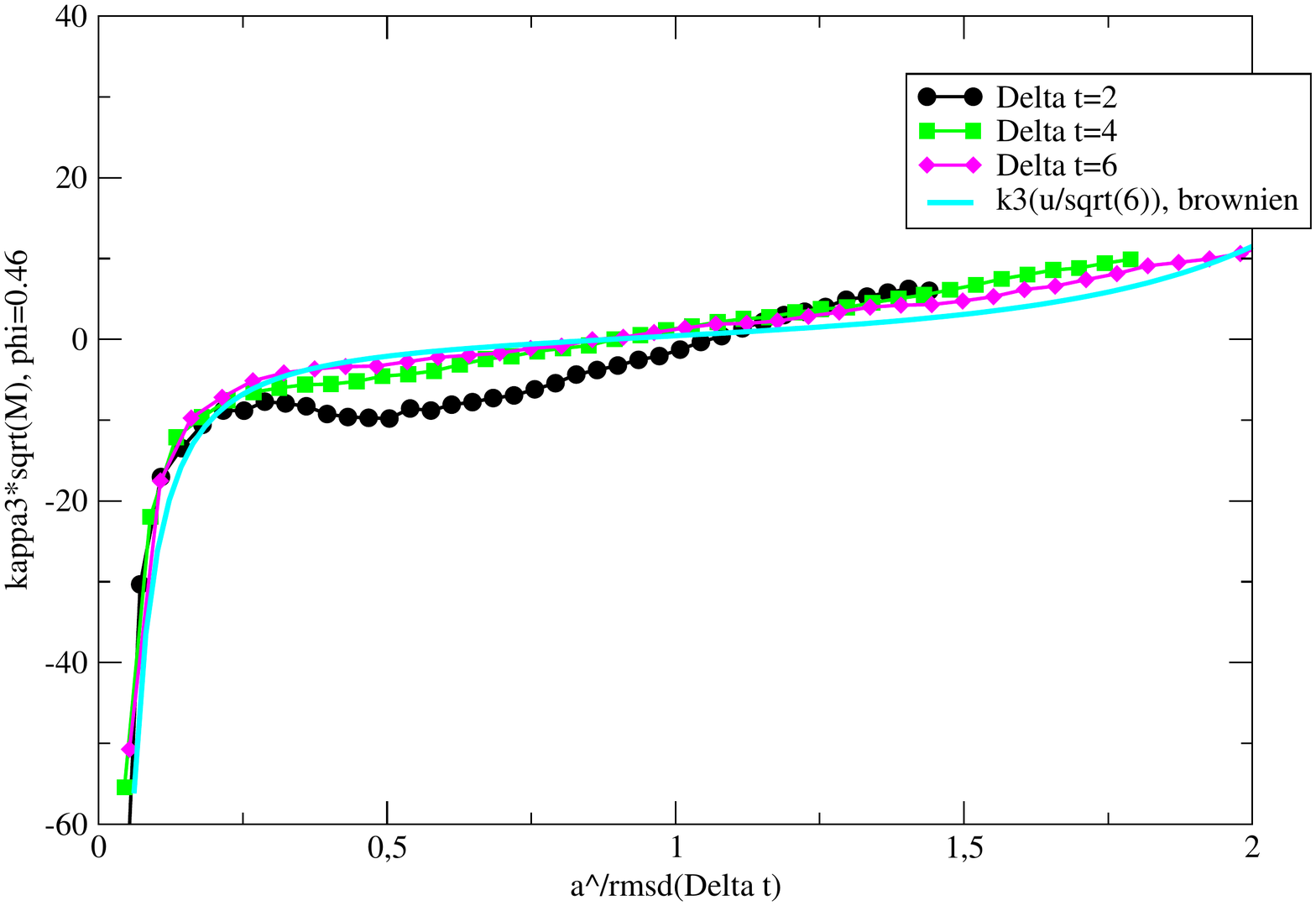}
\includegraphics[width=.49\columnwidth]{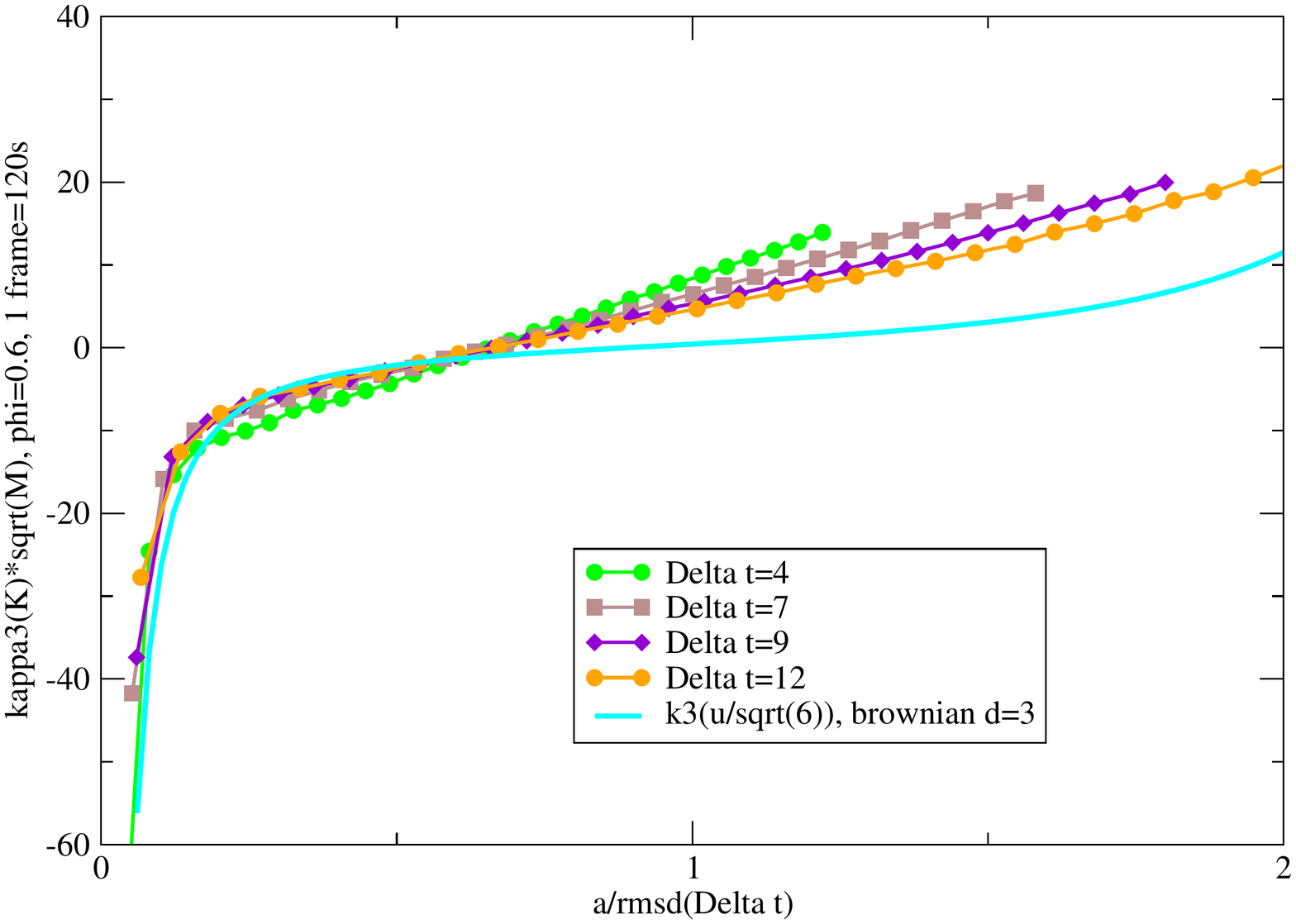}

\end{center}
\caption{
(\textbf{Right}) Scaled skewness $\kappa_3 \sqrt{M}$ at $\phi=0.6$ for varying values of $\Delta t$ in units of $\delta t_{0.46}$.
(\textbf{Left}) Scaled skewness $\kappa_3 \sqrt{M}$ at $\phi=0.46$ for varying values of $\Delta t$ in units of $\delta t_{0.6}$.
}
   \label{fig:WeeksSkew0.46_0.6}
\end{figure}

In Fig.~\ref{fig:WeeksSkew0.46_0.6}, %, Fig.~\ref{fig:WeeksSkew0.46-0.52-0.6} 
%and  Fig.~\ref{fig:WeeksSkew0.6}, 
we illustrate (as for our experimental data in~Fig.~\ref{skewRemy1} and~Fig.~\ref{skewRemy2}) that the more glassy (denser) system displays a strongly asymmetric skewness compared to the Brownian case.
For $\phi=0.46$, the skewness is very close to the Brownian case and close to zero for a large range of values of $u$, which is consistent with the symmetric histograms plotted in Fig.~\ref{fig:WeeksHisto0.46}.
For $\phi=0.6$, the skewness is also the same as the Brownian case for small values of $u$ but departs strongly from this behavior for $u$ larger than 0.6. This is again consistent with the symmetries revealed in Fig.~\ref{fig:WeeksHisto0.6}.
In the same way we did for our experiment on pNipam, we interpret the excess of larger than average activity for large $u$ as the manifestation of long range collective rearrangements (cooperatively rearranging regions) in glassier systems. However we observe that  the skewness for small $u$ is the same as in the Brownian case, hence one cannot see in these data an excess of less than average active events at small scales.
This may be due to the differences in scales probed by the two experimental setups. In Weeks's experiments, small $u$ corresponds to rattling inside a cage at very small length scales, whereas large $u$ corresponds to escapes from the cages. In our experiments the same difference in scales is true but a single movie will probe more escapes from cages, and less rattling motions, than in Weeks' setup. We believe that if the timescale of Weeks' experiments were comparable to ours, the departure from Brownian at negative skewness would also be visible.

\bibliographystyle{apsrev4-1.bst}
\bibliography{fredbiblio}

\end{document}